%% file: ms.tex
\documentclass[12pt,preprint]{aastex}

\def\pc{{\rm\,pc}}

\def\kyr{{\rm\,kyr}}
\def\Myr{{\rm\,Myr}}

\def\AU{{\rm\,AU}}

\def\erf{{\rm\,erf}}
\newcommand{\dgrs}{\ensuremath{^\circ}}

\lefthead{Berukoff \& Hansen}
\righthead{Core Dynamics}

\begin{document}

\title{Cluster Core Dynamics in the Galactic Center}

\author{Steven J. Berukoff and Bradley M.S. Hansen}
\affil{Department of Physics and Astronomy, University of California,
    Los Angeles, CA}

\begin{abstract}
We present results of N-body simulations aimed at understanding the dynamics of young stars near the Galactic center.  Specifically, we model the inspiral of a cluster core containing an intermediate mass black hole and $N \sim 50$ cluster stars in the gravitational potential of a supermassive black hole.  We first study the elliptic three-body problem to isolate issues of tidal stripping and subsequent scattering, followed by full N-body simulations to treat the internal dynamics consistently.  We find that our simulations reproduce several dynamical features of the observed population.  These include the observed inner edge of the claimed clockwise disk, as well as the thickness of said disk.  We find that high density clumps, such as that claimed for IRS13E, also result generically from our simulations.  However, not all features of the observations are reproduced.  In particular, the surface density profile of the simulated disk scales as $\Sigma \propto r^{-0.75}$, which is considerably shallower than that observed.  Further, at no point is any significant counter-rotating population formed.
\end{abstract}


\section{Introduction}
In the past decade, observations have conclusively established the presence of a supermassive black hole (SMBH) at the Galactic Center (GC) (Sanders 1992, Haller et al. 1996, Ghez et al. 2003).  Efforts to measure the mass of the SMBH have led to a large body of evidence regarding the stellar kinematics and mass distribution in this region (Genzel et al. 2003, Ghez et al. 2003, Sch\"odel et al. 2003, Paumard et al. 2005, Maillard et al. 2004, Eisenhauer et al. 2005, Ghez et al. 2005).  The most surprising findings include the discovery of a number of young, massive stars closely orbiting the central object Sgr A*, young stellar populations in two possibly counter-rotating stellar disks orbiting Sgr A* further out, and small stellar clumps or associations (so-called ``comoving groups'') within these disks (Levin \& Beloborodov 2003, Genzel et al. 2003, Lu et al. 2005, Maillard et al. 2004, Sch\"odel et al. 2005). 

The origin and peculiar kinematics of these structures beg theoretical explanation. Formation of stars near the SMBH is problematic because the strong tidal field is likely to shear and disrupt normal molecular clouds well before they can gravitationally collapse.  Other formation scenarios include molecular cloud collisions, star-star collisions, and giants whose envelopes have been tidally stripped, but none of these is entirely satisfactory. Current opinion falls into one of two classes -- either formation of stars by gravitational instability in an AGN-like accretion disk (Kolykhalov \& Sunyaev 1980, Shlosman \& Begelman 1989, Morris 1996, Sanders 1998, Goodman 2003, Levin \& Beloborodov 2003, Nayakshin \& Cuadra 2005) or rapid inward transport (due to dynamical friction) and subsequent tidal disruption of a star cluster that formed at larger radii (Gerhard et al. 2001).  Early numerical simulations of the latter scenario revealed that the cluster would not survive the infall to small radii unless extraordinary demands were placed on the mass and stellar density (McMillan \& Portegies Zwart 2003, Kim \& Morris 2003). An enhancement of this idea was the inclusion of an intermediate mass black hole (IMBH) at the center of the cluster, which served to both maintain the cluster potential well and slow internal relaxation (Hansen \& Milosavljevi\a'c 2003).  Further direct simulations, including this refinement, again placed stringent demands on the cluster initial conditions, although the required core density decreased (Kim, Figer, \& Morris 2004).  Recent Monte Carlo and N-body simulations of this process have demonstrated that $\sim 100$ stars can be transported via dynamical friction to about $1\pc$ from the GC, with an IMBH formed naturally through a runaway process of stellar merger during the migration (Baumgardt et al. 2004, G\"urkan \& Rasio 2005).  However, these simulations could not accurately follow the further evolution of this cluster core because of algorithmic limitations.  It is the principal goal of this paper to follow the physics of this process further inwards to determine to what extent this scenario may reproduce the observed features of the young star distribution.

To date, the reliability of existing simulations has effectively ended at about $1\pc$.  A chief cause of failure of these codes is the presence of very strong tidal fields due to the nearby SMBH, which causes normally simple algorithms for energy conservation and treatment of close encounters to become delicate and highly complex, resulting in occasional failure.  Since the majority of the interesting and puzzling observations have been made interior to this radius, simulations that might illuminate answers to these riddles are necessary.   

This paper discusses simulations of the dynamics of remnant cluster cores as they sink towards the GC,  specifically focusing on the region interior to $1\pc$.  The paper is organized as follows:  \S~\ref{sec:nummeth} describes the numerical methods of simulating the inspiral of general, three-body and $\sim 50-$body systems, including the implementation of dynamical friction that creates the inspiral.  \S~\ref{sec:3body} describes the three-body simulations, and \S~\ref{sec:nbody} the N-body simulations, and compares and contrasts the two regimes.  Finally \S~\ref{sec:disc} concludes with a discussion of how these results can be used to understand dynamics at the Galactic Center, with particular regard to the curious observed structures such as the S-stars and the comoving groups IRS13E and IRS16SW.

\section{Numerics}
\label{sec:nummeth}
\subsection{Overview}
While a large body of literature exists covering many aspects of N-body dynamics, few numerical studies incorporate strong tidal fields.  Strong interactions between stars, when handled improperly, lead to large energy errors and subsequent spurious results, and strong tidal fields encourage stronger interactions.  The two canonical methods of dealing with close encounters in N-body integrations are softening and regularization.  By the inclusion of a small softening parameter into the denominator of the force calculation, the effects of small interparticle separations can be avoided, at the cost of reduced accuracy for close encounters.  In simulations for which the particle density or interaction cross-section are relatively low, such as large-scale galaxy simulations, this method is useful and often employed.  In this paper, the focus is on both individual dynamics and in statistical averages of these dynamics in dense environments; therefore, softening is inappropriate, as some portion of the essential motions would be lost.

\subsection{Regularization}
\label{subsec:Regular}
Regularization is a technique in which a close encounter between particles with $1/R$-style force terms is mathematically transformed into a single center-of-mass particle, the singularity removed, and its path integrated for a suitable time (see, e.g., Aarseth 2003).  The essential benefit of this method is that the close encounter is correctly integrated with minimal error, and thus the individual dynamics are more realistic.  Several schemes are in widespread use; two of the most common are two-body, or Kustaanheimo-Stiefel (KS), regularization, and ``chain'', which is essentially a coupled set of KS regularizations for multiple particles, in which each pair of neighbors is KS-regularized in a link-list fashion, but without closure of the chain.  We utilize the benefits of these approaches, although for the systems under consideration here they are not without pitfalls.

KS regularization is analytically straightforward, and its extension to chain regularization is more complex but tractable.  Typically, though these algorithms are well-developed, they are optimized for similar-mass particles in the absence of tidal fields.  When faced with large mass ratios, the decision-making behind the initiation, continuation, and termination of regularization can be poorly defined, leading to frequent integration errors.  The addition of strong tidal fields further complicates matters, as the regularized pair becomes subject to a perturbation that can be of the same order-of-magnitude as the two-body interaction itself.  These problems may be alleviated by better decision making, including that governing  regularization and the treatment of high-velocity intruders.  However, the central difficulty is a basic inadequacy of the algorithms, and no amount of tweaking is going to produce a universally successful numerical treatment.

A common configuration in these simulations consists of several particles that orbit close to the massive particle (``IMBH'').  Each particle-IMBH pair requires regularization to correctly compute the two-body orbits.  Simultaneously, all close particle-particle pairs need the same treatment.  This type of configuration is not properly dealt with by modern chain regularization, the only fully implemented multiple regularization technique commonly available.  There is an alternative algorithm, called ``wheel-spoke'' (Aarseth 2003), which could remedy this problem, but it is not yet mature and currently suffers from a number of fatal setbacks. (Aarseth, private communication)  

This inadequacy of current regularization techniques somewhat limits the applicability of our present work.  As the IMBH inspirals and attempts to transport multiple tightly bound stars down the potential well, the present algorithms must continually switch between regularizing the IMBH-particle and particle-particle pairs.  This is inefficient and can be error-prone; for example, when two or more particles are each in hard binaries with the IMBH, the integration steps may be incorrectly calculated, resulting in erroneous orbits.  The termination of one of these regularizations can result in the introduction of incorrect orbits into the calculation.  Alternatively, drastic changes in the orbits may occur, causing high velocity ejections in directions normal to their orbital planes.  Indeed, this was observed in some simulations, but further analysis showed that these events were caused by numerical error rather than basic physics.  So, while one might be tempted to excitement at the observation of high-velocity ejections in the simulations, such results are not always correct.  Given the systems simulated here, such errors place computational lower limits on the possible simultaneous star-IMBH distances, both during the simulation and when creating initial conditions, limiting the density of the initial stellar systems and consequently our ability to simulate strongly bound subsystems deep into the SMBH potential well.

In the present study, most of the N-body runs were recalculated at least once, and care was taken to ensure that situations arising from poor regularizations were removed from the final data analysis.  This was done in a variety of ways, including 
\begin{itemize}
\item the analysis of the energetics of difficult configurations;
\item verifying that the timesteps used in regularizations were appropriate to the system being regularized;
\item identifying and tracking the progress of multiple simultaneous regularizations; 
\item tuning parameters to avoid the unwarranted initiation of chain regularization, when appropriate.
\end{itemize}

Difficult dynamical configurations arose from a number of situations, caused sometimes by algorithmic difficulties.  For instance, consider again the common scenario of an IMBH bound to several close particles, with a high-velocity particle intruder.  This interloper is close enough to the IMBH to require regularization, but moving sufficiently fast that it might be in the region for only a short time.  The basic criterion for the initiation and termination of regularization requires only information about the relative separations of the particles.  In addition, the onset of a regularization period is controlled by several parameters that are set at the beginning of the simulation but that, on short timescales, have only a limited ability to adapt to the environment being integrated.  In this case, the intruder particle initiates a regularization because of its proximity, and significantly alters the regularization environment for the other stars near the IMBH.  This is unfortunate because the intruder moves off, leaving a computationally error-prone system behind.  Alternatively, the intruder does not trigger a regularization, but interacts strongly with several members of the core.  Its velocity is high, and the required timestep to properly treat the interactions is too long to maintain low energy error.  The regularization parameters are updated a short time later, but by then the intruder has gone on its way, leading again to an untenable configuration.  Such issues are typical in these types of simulations; their identification and analysis may indicate a path toward improved techniques.

\subsection{Particulars}
In all of this work, Aarseth's NBODY6 was employed for the direct integration of stellar orbits, with several modifications.  Besides needing the basic N-body integration engine, requirements for a drag term and regularization of close encounters in the presence of strong tidal fields placed strict constraints on the algorithms used, and, often, required some adjustment.  NBODY6 includes several subroutines which compute a tidal field due to a variety of scenarios, but none are well-suited to the extreme mass ratios and tidal fields considered here, and were not used.  Instead, new algorithms were built and integrated into the current experiments.  NBODY6 also includes modern KS and chain regularization techniques for minimizing error during close encounters between particles, but, as discussed above, these methods are tailored primarily to regularizing similar-mass encounters, and current implementations are not as agile when faced with large mass ratios.  Thus, constants in the vanilla NBODY6 which govern the classification of close encounters and the initiation and termination of regularization were monitored and tuned to maintain the integrity of the simulations.

Initial phase space coordinates were created using an isotropic King $W_{0}=9$ model generator.  The code uses a 1D Poisson solver and fifth-order Runge-Kutta integrator with adaptive step size.  The inclusion of a central black hole in the distribution is achieved by adding the potential of the black hole to that of the King model, then computing a new distribution function based on this potential and the original density profile.  Candidate sets of initial conditions are then selected based on minimizing the moment of inertia and its derivative, then allowing the cluster to partially virialize.  These clusters thus come as close to dynamical equilibrium as our simple model will allow.  The cluster core, or, in the three-body case, the IMBH and star, are then placed at $1\pc$ away from the SMBH on the x-axis, with an initial velocity appropriate for the eccentricity used.  Other specifications of initial conditions specific to either the three-body or N-body case are detailed in their respective sections.

We assume Chandrasekhar dynamical friction as a drag on the cluster, forcing inspiral.  This treatment must be handled carefully, for two reasons.  This approximation depends on the assumptions that the stars are uniformly distributed and their velocity distribution is isotropic, and in practice, the Chandrasekhar formula provides a reasonable estimate of the drag induced on an orbiter.  However, it fails to accurately describe strongly inhomogeneous systems in which the forces applied to the orbiter are nonuniform.  An extension of the standard paradigm, in which the Holtsmark distribution characteristic of the Chandrasekhar formalism is generalized, concluded that inhomogeneities drive a stronger drag against orbiting bodies than would be expected with a smooth distribution (Del Popolo \& Gambera 1999).  This is important in cases where the gravitational field is strongly discretized due, for example, to nonuniform stellar density.  During early experiments, however, we found that the inhomogeneities need to be very dense and localized (such as those from high-density molecular cloud cores) to produce significant changes, and so may be neglected here.

Second, the Hermite integration utilized in NBODY6 depends not only on the force, but also on its three derivatives, and the correct treatment of energy error relies on a reasonable estimate of the work performed by any putative drag.  The standard Chandrasekhar formula for dynamical friction is (Binney \& Tremaine 1987)
\begin{equation}
\frac{d {\bf v}}{dt} = -4 \pi \ln \Lambda G^2 M \chi \rho (r) \frac{{\bf v}}{v^3}
\end{equation}
for
\begin{equation}
\chi = \erf\biggl(\frac{v}{\sigma \sqrt{2}}\biggr) - \sqrt{\frac{2}{\pi}} e^{-v^2/2\sigma^2},
\label{eq:basicDF}
\end{equation}
$\ln \Lambda$ is the Coulomb logarithm, and $\rho (r)$ is the background stellar density which interacts with the moving body to induce drag.  In this paper, time derivatives of $\chi$ can be eliminated, since its growth is essentially adiabatic.  We found that this slight improvement provides a significant reduction in computational time without affecting the results.

Once stars are lost from the control of the IMBH, subsequent interactions occur in a disk.  A typical cluster is a spheroid with a halo+core structure, often created by mass segregation.  As it is tidally stripped by a SMBH, its former members are strewn into orbits around the SMBH, with initial orbital semimajor axes, eccentricities, and inclinations similar to that of the IMBH.  A large cluster that begins inspiral far from the depths of the SMBH potential well disrupts, leaving a wide concentric swirl of stellar debris in its wake.  Disrupted clusters with large initial populations ($\sim 10^5$ stars) thus create a density enhancement over and above that of the background.  This can amplify the deceleration due to the drag from the {\em background} density, given in Eq.~\ref{eq:basicDF}, resulting in a short infall time.  In these experiments, the tidal tail of the small cluster core has a much smaller stellar population.  Therefore, the modelled analytic background stellar population is the primary source of drag for the IMBH and its former cluster members. 
   
\section{Three-body}
\label{sec:3body}
\subsection{Setup}
\label{sec:3bsetup}

We start with the limited case of an IMBH with only a single star orbiting it.  Simulating the dynamics of such a system inspiralling into a massive potential well will isolate the physics of tidal stripping and subsequent mutual scattering. In following sections we compare with the case of multiple stars to understand the role of internal dynamical evolution in the cluster. The main causes of variation in the three-body results are different IMBH eccentricities and change of inspiral speed, while other parameters such as the presence of a mass spectrum proved to be unimportant, and little mention will be made of them.  Plots shown in the following sections are representative of results obtained, and do not contain data from the full 10000 runs, unless specified.

In order to more fully understand the relevant parameter space for the cluster core simulations, a large suite of three-body runs is performed.  Initial conditions are created by generating King models, as described above, each with 1000 particles, with individual stars randomly selected from the phase-space distribution and placed into a three-body system with an IMBH and an SMBH point-mass potential (with $M\sim 4\times10^6 M_{\odot}$), whose motion is not integrated.  The simulations employ a rough $M_{imbh}:M_{star}$ mass ratio of $10^3:1$, and a range of IMBH masses are used (250, 500, 750, 1000, 1200, 1500, 2000 $M_{\odot}$).  Stars whose initial separation from the IMBH are more than their Jacobi radii are rejected.  Selected stars are assigned a mass from a Kroupa-type (Kroupa, Tout, \& Gilmore 1993) initial mass function, slightly modified to include stars of up to $100 M_{\odot}$.  The IMBH+star system is then placed on an initial orbit with one of four eccentricities: $0$, $0.2$, $0.5$, $0.8$.  For each value of the IMBH mass and initial IMBH eccentricity, $300$ stars are selected and simulated.

In order to force an inspiral, a drag term is applied to the IMBH, in the form of Eq. (\ref{eq:basicDF}) with $\sigma = v/\sqrt{2}$, which is roughly an isothermal distribution, or $\rho \propto r^{-2}$.  For these runs, the detailed density structure of the Galactic Center environment is unimportant, as the primary goal is to understand what effect star-star interactions have on cluster evolution by contrasting the three-body case with the more general N-body case.  A basic inspiral lasts approximately $15 \Myr$, although faster ($\sim 5 \Myr$) and slower ($\sim 100 \Myr$) inspirals are also tested.  This is a rough range of inspiral timescales for circular orbits, which clearly represent an upper limit when performing eccentric inspirals.  (McMillan \& Portegies Zwart 2003)

Simulations are terminated under one of two conditions: the IMBH migrated to within $0.01 \pc$ of the origin (SMBH); or the simulation fails, typically due to the ejection of a high-velocity particle (see \S~\ref{subsec:Regular}) or the formation of a very hard star-IMBH-star triple with semimajor axes $\sim 20\AU$.  Approximately 10000 runs were conducted using a dual-processor Linux workstation.  

\subsection{Results}
\label{sec:3bresults}

\begin{figure}
\plotone{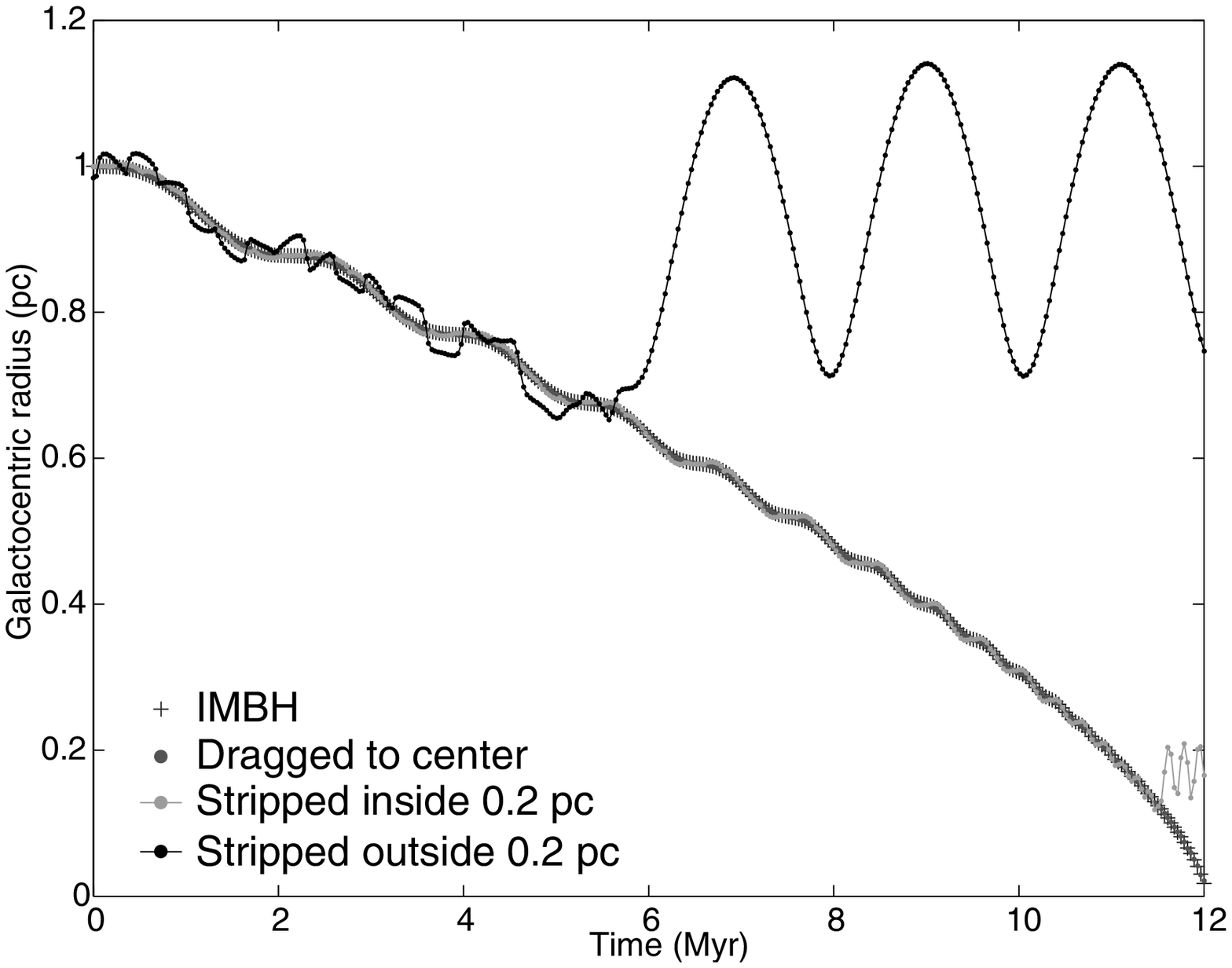}
\figcaption{
The ability of the IMBH to transport stars along its inspiral is clearly demonstrated.  The orbits of the IMBH and one starcontinuously shrink until the end of the simulation, remaining bound.  One star is stripped inside 0.2 pc, and one star survives little of the inspiral.  Generally, stars that are stripped obtain orbital elements ($a, e, i$) that do not differ greatly from that of the IMBH, although there is a large scatter in this distribution.  See also Figures~\ref{fig:3b_ecc} and ~\ref{fig:3b_inc}.
\label{fig:dragging}}
\end{figure}

The presence of an IMBH provides a necessary and sufficient mechanism for transporting a star down the potential well to a tidal stripping radius.  Figure~\ref{fig:dragging} shows different stars that are transported partially or fully toward the SMBH.  The majority of stars are stripped and typically end up on large semimajor-axis orbits.  However, a few stars are transported deep into the well, with those most tightly bound dragged to the simulation boundary.  Just what fraction survive will be quantified below.

\subsubsection{Effect of IMBH Eccentricity}
\begin{figure}
\plotone{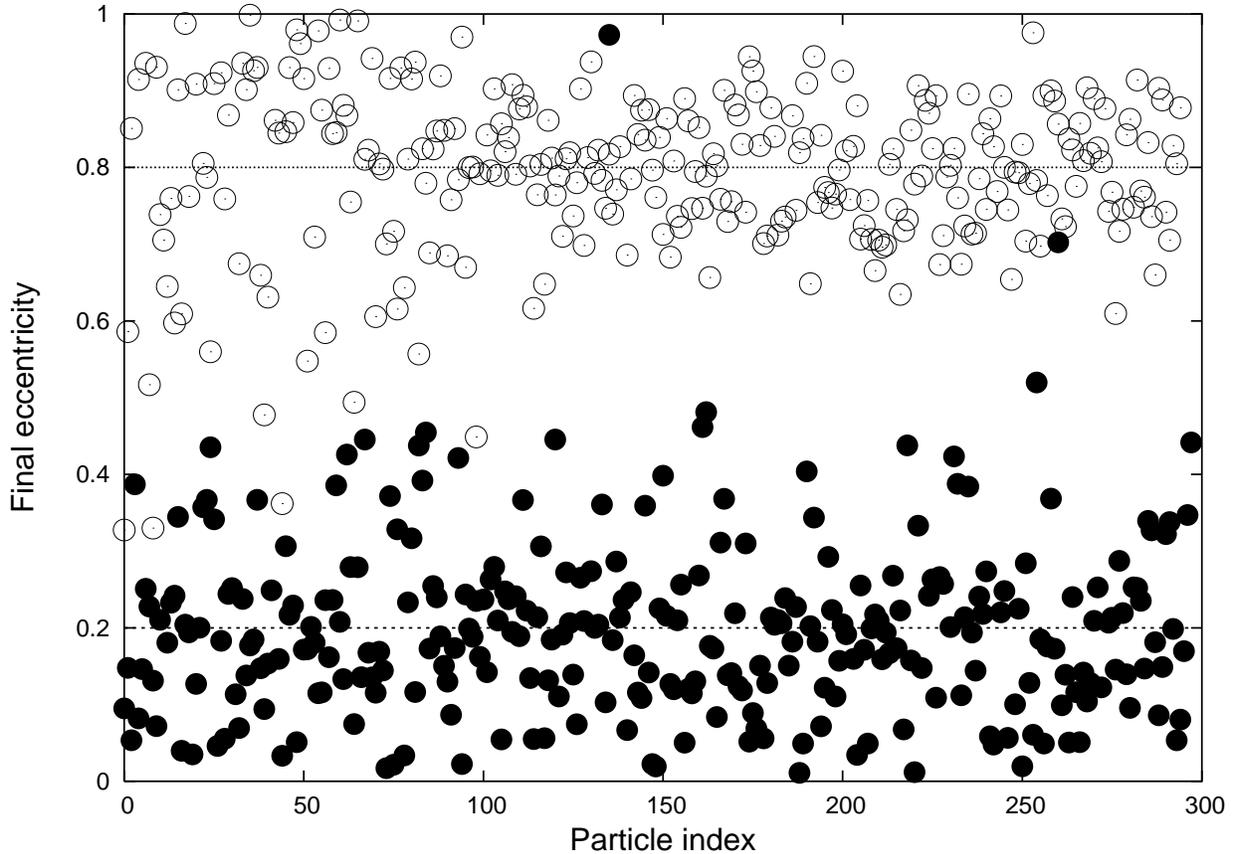}
\figcaption{Final stellar eccentricities mirror that of the IMBH.  The lines represent the initial IMBH eccentricity.  There is a large scatter about each line, but the only rarely do stars acquire eccentricities in their orbits about the SMBH that differ largely from that of the IMBH.  Note that there are a few stars that began as part of the $e=0.8$ group that are scattered into more circular orbits, but few stars initially with $e=0.2$ orbits scatter into much higher eccentricity orbits.
\label{fig:3b_ecc}
}
\end{figure}

For a given IMBH initial eccentricity, the final stellar eccentricities tend to somewhat mirror that of the IMBH.  This is not surprising, since the cluster's systemic velocity is larger than the orbital velocities of stars about the IMBH when they are near the Roche lobe.  Figure~\ref{fig:3b_ecc} shows this relationship for initial IMBH eccentricities of 0.2 and 0.8.  Note that there is significant scatter about the IMBH value, for both cases; further, note that the scatter for some stars in the $e=0.8$ case is larger, due to a strong perturbation during a highly eccentric encounter. 

\begin{figure}
\plotone{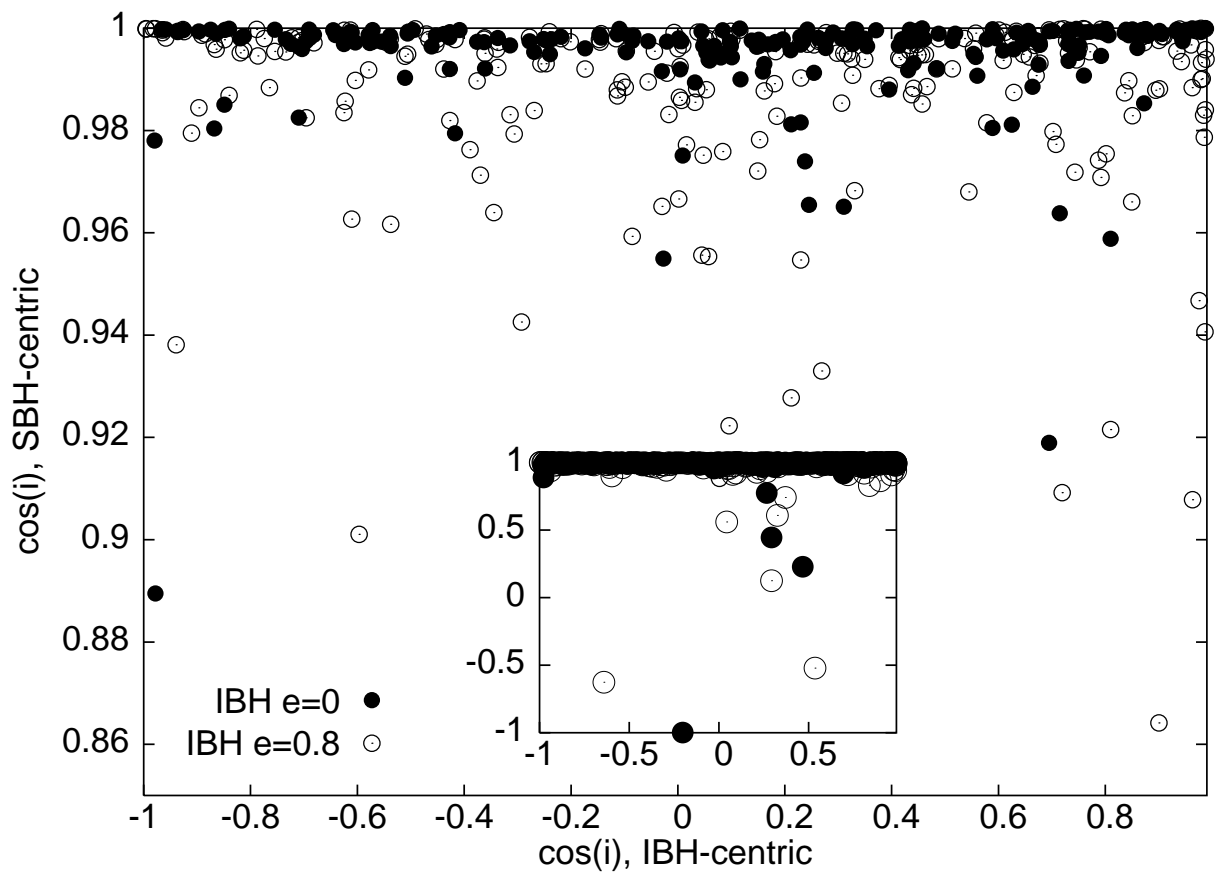}
\figcaption{
Stars initially on low inclination IMBH orbits tend to stay there, while those with high inclination are likely to be perturbed into low inclination orbits, reflecting their deposition into the IMBH orbital plane.  The inset shows a small number of stars that differ from this trend.  Of these, only $\sim 0.01-0.1\%$ acquire retrograde, or counter-rotating orbits.
\label{fig:3b_inc}
}
\end{figure}

Similarly, Figure~\ref{fig:3b_inc} shows this behavior for stellar inclinations.  Stars beginning the simulation bound to the IMBH in its orbital plane have initial inclinations that trend linearly with their final states.   Stars orbiting the IMBH far from its orbital plane are mapped to small final inclinations.  Again, since the concominant motion is that of the IMBH, these results are not surprising.  Of particular interest, however, are any stars that end with high inclinations, possibly rotating in a retrograde fashion, which would help to explain the source of the putative counter-rotating disk at the Galactic Center.  Note from Figure~\ref{fig:3b_inc} that there are a small number of stars with such orbits; statistically, summing over all runs, these account for approximately $0.01-0.1\%$ of the final states.  For a remnant core of a globular cluster, retaining perhaps $10^3$ members, this results in perhaps a handful of stars in such orbits, assuming multiple stars could be transported in this manner, a subject discussed in the context of the N-body runs below.

\begin{figure}
\plotone{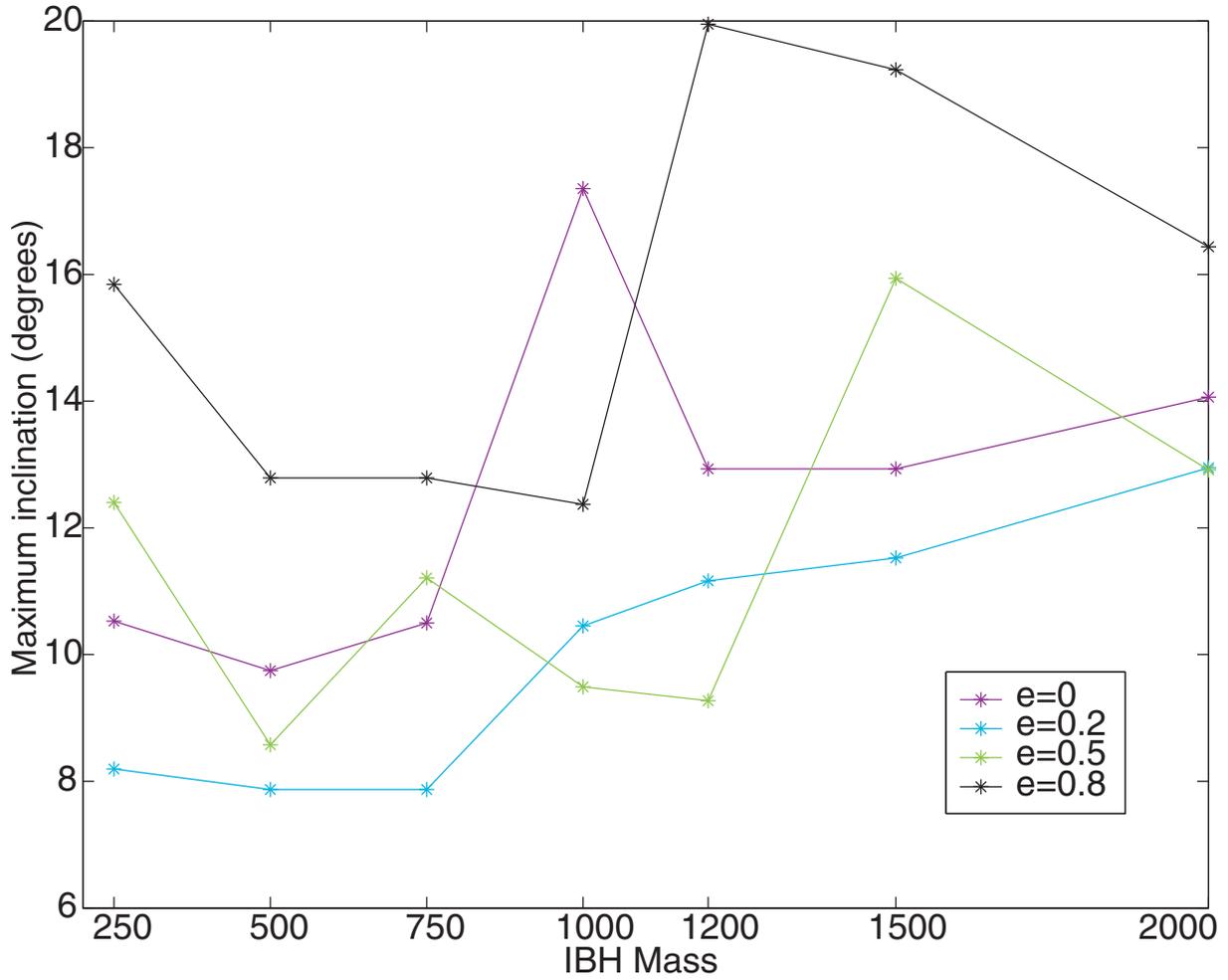}
\figcaption{
Marginal trend toward increased average stellar inclination as IMBH mass increases.  This suggests that the larger IMBH mass promotes stronger interactions during stripping.
\label{fig:3b_inc_e}
}
\end{figure}

There is a slight trend toward increased average stellar inclination as the initial IMBH eccentricity increases, for all IMBH masses simulated.  Figure~\ref{fig:3b_inc_e} reflects this; the difference is not large, but is clearly discernible.  The likely culprits are strong encounters near the IMBH peribothron.  The IMBH orbital plane can be viewed as a midplane for a disk composed of the orbits of stripped stars.  Since the stellar inclinations are nonzero, there is some scatter about the midplane.  Should a star be unlucky enough to find itself near the IMBH peribothron when the IMBH itself passes by, the scattering interaction may cause the star's inclination to grow due to the normal component of the perturbation.  Few of these interactions typically occur in a typical simulation, but their result is a small number of high-inclination stellar orbits. 

\subsubsection{Inspiral Speed}
Three inspiral speeds were used to understand the effect on final stellar distributions: ``regular'', corresponding to approximately $t_{insp}\sim 15 \Myr$, ``slow'' corresponding to $t_{insp}\sim 100 \Myr$, and ``fast'', with  $t_{insp}\sim 5\Myr$.  This was effected by changing the value of the constants slightly in Eq. \ref{eq:basicDF}.  Slow inspirals transport fewer stars to orbits with semimajor axes of less than $0.5\pc$, while fast inspirals leave fewer stars with semimajor axis greater than $1\pc$.  Slow inspirals promote larger final semimajor axes generally, for a relatively simple reason.  In a slow inspiral, a stripping event leaves the IMBH and star in similar orbits.  Subsequent IMBH passages can lead to significant scattering events.  Note also that the endpoint of a resonant trapping event (discussed below) can produce configurations where the IMBH and star orbits nearly coincide.  Since subsequent scattering is generally rare, the occurrence of strong star-IMBH interactions soon after stripping provides some estimate about the frequency of resonant trapping events.

\subsubsection{Miscellany}
\begin{figure}
\plotone{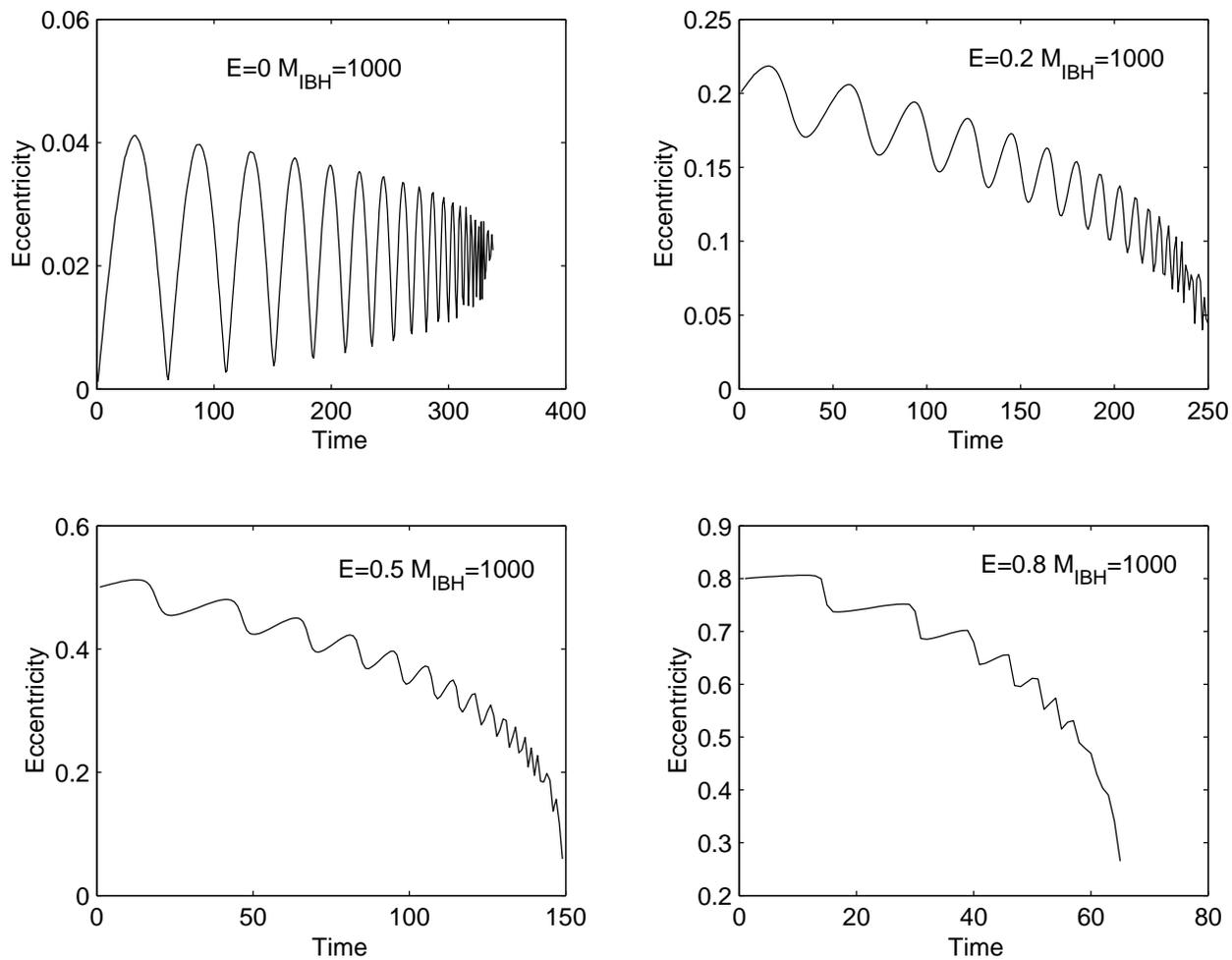}
\figcaption{
Circularization of IMBH orbits under dynamical friction.  Note that for high eccentricity orbits, damping is strong, whereas for low eccentricity it is the oscillations that are damped.
\label{fig:dfc}
}
\end{figure}

Generally $5-10\%$ of stars arrive in the inner $0.2\pc$, with the stellar orbits determined primarily by the initial IMBH eccentricity and the speed of inspiral.  There are other effects, however, that are of perhaps anecdotal interest that we mention here.  Due to our implementation of dynamical friction for this three-body case, the IMBH orbits tend to circularize during inspiral, as seen in Fig.~\ref{fig:dfc}.  The circularization occurs for all initial IMBH eccentricities, and is caused by the tangential form of the drag.  Gould \& Quillen (2003) show that in a Kepler potential, a drag force due to a background density $\rho \sim r^{-\nu }$, will tend to circularize a stellar orbit if $\nu > 3/2$.  Recalling that our drag $\rho\sim r^{-2}$, this behavior is expected.  This is to be contrasted with realistic density profiles of the Galactic center used in the cluster core experiments below.

\begin{figure}
\epsscale{1.0}
\plotone{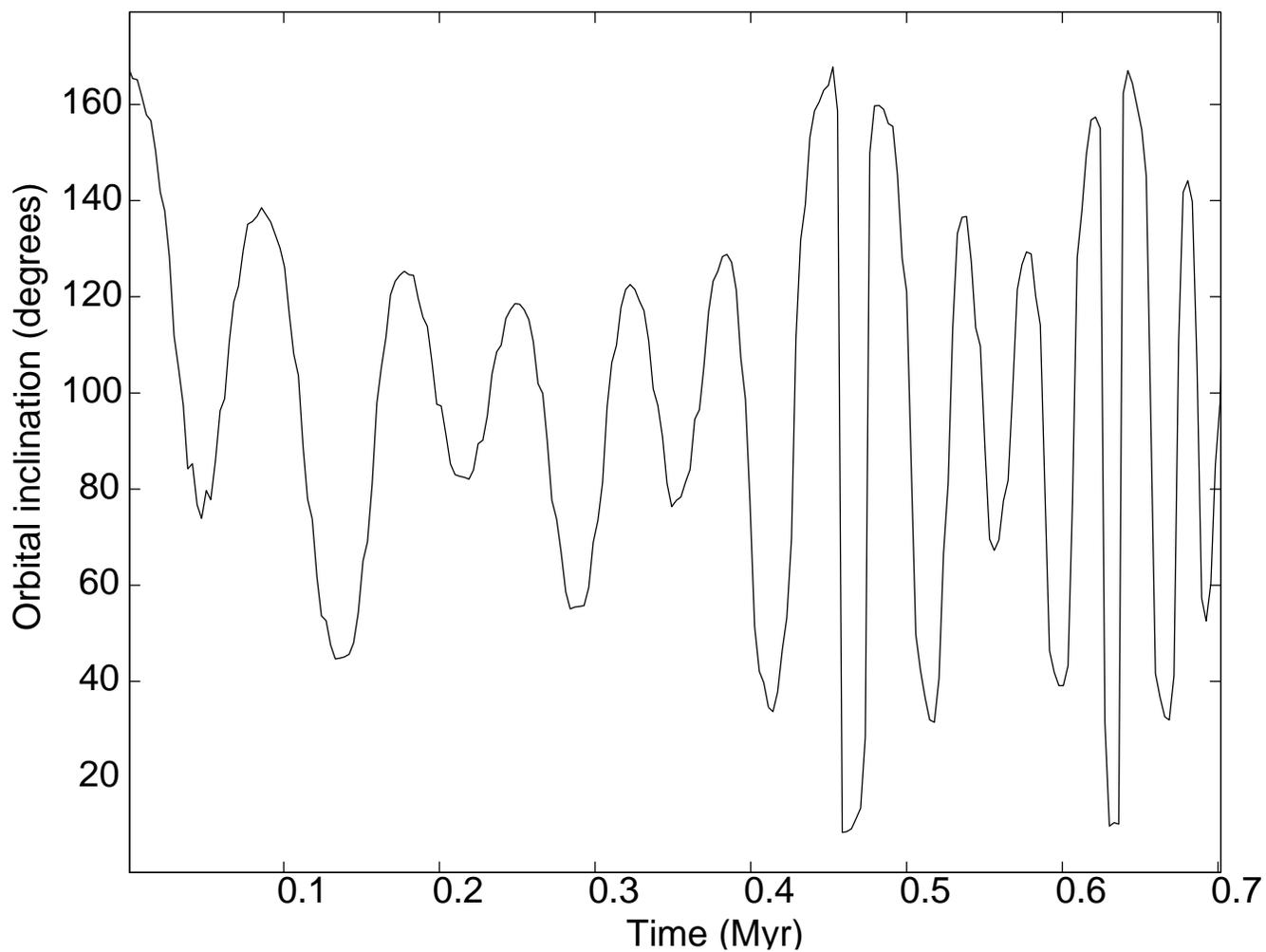}
\figcaption{
Large oscillations in stellar inclination, for a star still bound to the IMBH and therefore not yet stripped.  This is not necessarily indicative of Kozai-type resonances being induced in the star, but rather is a consequence of the nonlinearities induced in the stellar orbit due to the orbital motion of the IMBH as it migrates inward.
\label{fig:incp}
}
\end{figure}

Extremely rich dynamics are often lost in studying dynamical systems statistically.  For instance, for a star that has yet to be stripped, its motion around the IMBH is highly non-Keplerian owing to the IMBH's motion about the SMBH.  This can have profound influences on the star's orbital elements, causing large oscillations in the inclination (relative to the IMBH orbital plane) as seen in Figure~\ref{fig:incp}.  Once the star is stripped, such phenomena are extremely rare, but prior to stripping, this behavior is fairly common for stars beginning on large inclination orbits.  

Another phenomenon observed in the simulations is resonant trapping (Murray \& Dermott 2003).  Immediately after a star gets stripped, it can wander near the IMBH's $L_4$ and $L_5$ Lagrange points.  While there, the star's orbit is now SMBH-centric, but its orbital elements mirror that of the IMBH.  This will not last for long, as the Lagrange points are unstable due to the drag force causing inspiral.  However, the star will remain near the Lagrange point for perhaps a few tens to hundreds of years.  After the star has wandered too far from the equilibrium points, its orbit will still be roughly that of the IMBH; for slow inspirals, this can cause strong subsequent star-IMBH encounters and change the orbital elements significantly.  

Observations of individual particle orbits indicate that resonant trapping occurs in perhaps $1\%$ of the orbits.  Resonant trapping can be identified by a combination of a Roche-style criterion and an escape condition: when the escape condition is fulfilled but the Roche condition is not, a flag may be raised and a more detailed examination of the dynamics is undertaken.  Often it is the case that when both conditions are satisfied, the star is no longer trapped, although this is not foolproof, and merits further work.  Resonant trapping could be of importance in future studies examining the detailed microdynamics of the stripping process, or of the dynamics of few-body exchange, which has direct relevance to stars' inheritance of orbital parameters from the IMBH.

\section{N-body ($N \sim 50$)}
\label{sec:nbody}
\subsection{Setup}
\label{sec:nbsetup}
The basic initial setup has previously been described in \S~\ref{sec:nummeth}, and for these N-body runs, a number of additional features apply.  The cluster cores were generated with virial radius $R_{vir}~\sim 0.06\pc$, so that the outermost edge of the core lies slightly interior to the maximum Jacobi radius of any cluster stars.  The cluster is initially non-rotating, although a few rotating models were tested, yielding no significant differences.  Termination conditions were also similar, although an added condition was the formation of difficult hierarchies including two or more very hard binaries with the IMBH.  

In all, $211$ simulations were performed, with most models employing $\sim 50$ particles, although several simulations with $100$ and $200$ stars were tested as well.  These larger data sets produced no differences.  Multiple sets of initial phase space coordinates were used, and for each set, multiple masses of the IMBH were used, varying between $1015-5540 M_{\odot}$.  Further, for each IMBH mass, three initial eccentricities of $0$, $0.25$, and $0.5$ were used.  The SMBH mass was fixed at $4\times 10^6 M_{\odot}$.

All stellar masses were equal, the value being determined by a common set of simulation parameters, and typically ranged between $3-8$ solar masses, depending on the parameters used.  There are two complimentary reasons why the equal mass case is sufficient here.  First, in the absence of a tidal field, the mass segregation timescale for a group of stars this small is very short, on order $5\kyr$, which is an order of magnitude smaller than the orbital period at $1\pc$.  Thus, one would expect that such a core is always dominated by high mass stars.  The second reason is that the inspiralling cluster scenario posits that a cluster is delivered to the Galactic center after significant mass segregation has occurred in the weaker tidal field far from the GC (see, e.g., McMillan \& Portegies Zwart 2003).  By that time, the less massive stars, relegated to the halo, have been stripped from the cluster, and those stars that are delivered to the inner parsec are high-mass, and rather uniformly so.  Either way, the present simulations have an outer boundary of $1\pc$, and stars that begin their final descent are all be of roughly equal mass.  Approximately $50$ early experiments incorporating a mass distribution revealed no differences in the final stellar orbital elements or the creation of the various observed structures.

Two different schemes were used for the background density profile.  In the first, the density employed is from Genzel et al. (2003), where the detected stellar background is a broken power-law,
\begin{equation}
\label{eq:genz}
\rho (r) = 1.2 \times 10^6 \biggl(\frac{R}{0.4\pc}\biggr)^{-\alpha}  M_{\odot} \pc^{-3}
\end{equation}
where $\alpha$ is $2$ outside $0.4\pc$ and $1.4$ inside this radius.  This estimate is based solely on the detected stellar population, and does not account for any dark component.  Thus, in the second set of experiments, a putative dark population was added to Eq.~\ref{eq:genz}, with density 
\begin{equation}
\rho (r) = 1.68 \times 10^5 \biggl(\frac{R}{0.7\pc}\biggr)^{-1.75}  M_{\odot} \pc^{-3}
\end{equation}
which is a Bahcall-Wolf cusp of stellar-mass black holes (Bahcall \& Wolf 1976; Miralda-Escude \& Gould 2000).  In this paper, the addition of a dark-component cusp tests whether there are any significant effects on the properties of the final orbits of the former cluster members.

\subsection{Results}
\label{sec:nbresults}
In reporting the N-body outcomes, plots are representative of the results, since they are not statistically divergent between runs, with mention of deviations made when appropriate.

\subsubsection{Effect of Cusp}
The density enhancement introduced by the presence of the dark cusp has two significant effects on the global inspiral parameters, because its presence or absence affects the inspiral of the IMBH, which in turn dominates the local star-IMBH dynamics.  The first is the eccentricity change of the IMBH inspiral, which, if large and positive, would alter any stellar orbits in its vicinity.  From analytical estimates, the eccentricity can be expected to decrease if $\nu>3/2$ and increase for $\nu<3/2$ (Gould \& Quillen (2003)).  Outside of $0.4\pc$, the density profile is isothermal, and therefore the IMBH orbit will tend to circularize.  After passing through the knee, the density profile is shallower, but only slightly, than the threshhold cited above.  The presence of the cusp reduces the eccentricity growth of the orbit.  The density enhancement due to the cusp is more important inside $\sim 0.02\pc$, which is near the simulation end boundary.  

The second effect is much more significant.  The presence of the mass distribution of the cusp significantly decreases the inspiral time.  For initially circular orbits, the decrease is nearly a factor of $2$.  For initially eccentric orbits, the decrease is nearly a factor of $3$; the effect is stronger because eccentric orbits experience a stronger drag during pericenter passage, losing more energy to their surroundings.  The implications of this on how and why the massive OB stars can migrate through a variety of mechanisms to the Galactic Center are discussed in more detail in \S~\ref{sec:disc}.  However, besides these two effects, there was no statistically significant difference in the results between the simulations with a cusp and without.  

\subsubsection{Stellar Transport \& Comoving groups}
\label{sec:trans}
\begin{figure*}
\begin{center}
\begin{tabular}{cc}
\resizebox{80mm}{!}{\plotone{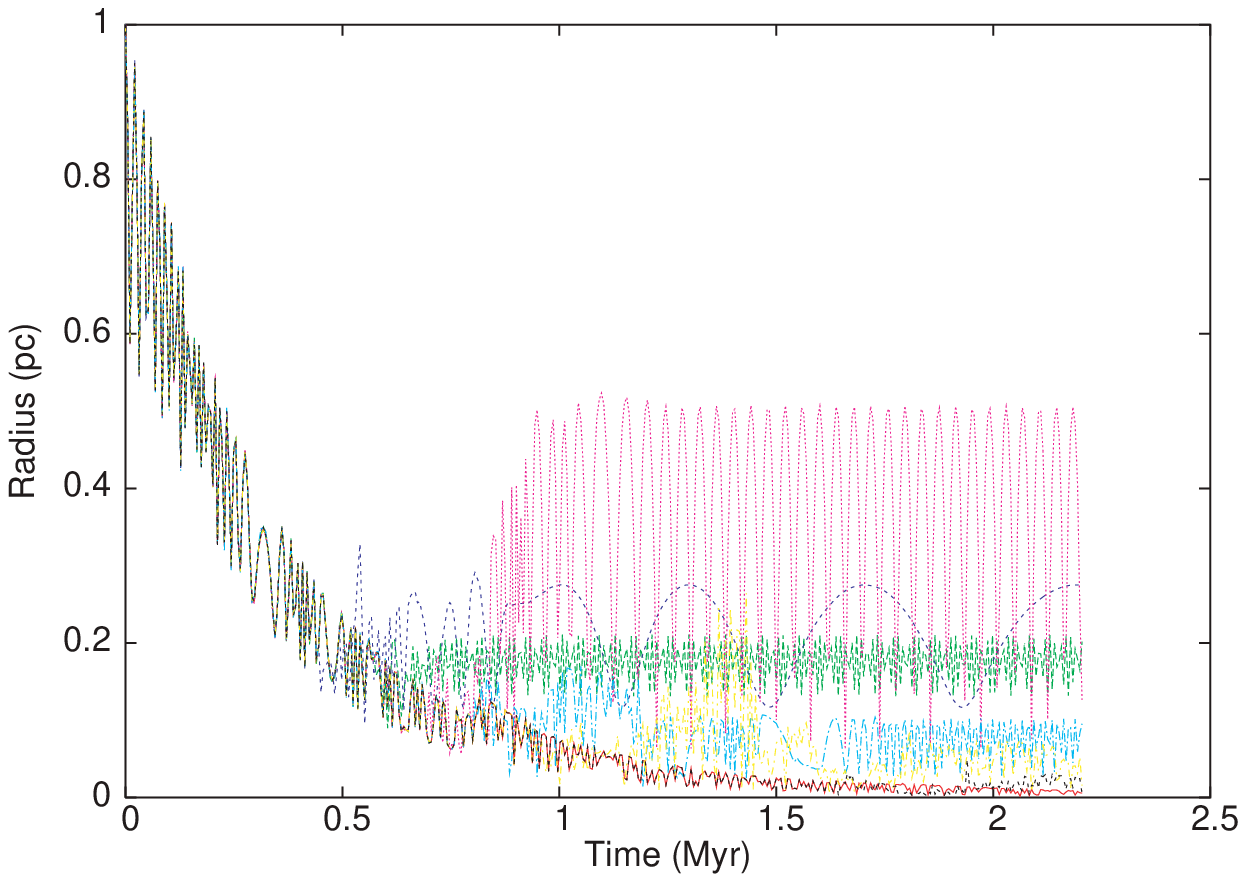}} &
\resizebox{80mm}{!}{\plotone{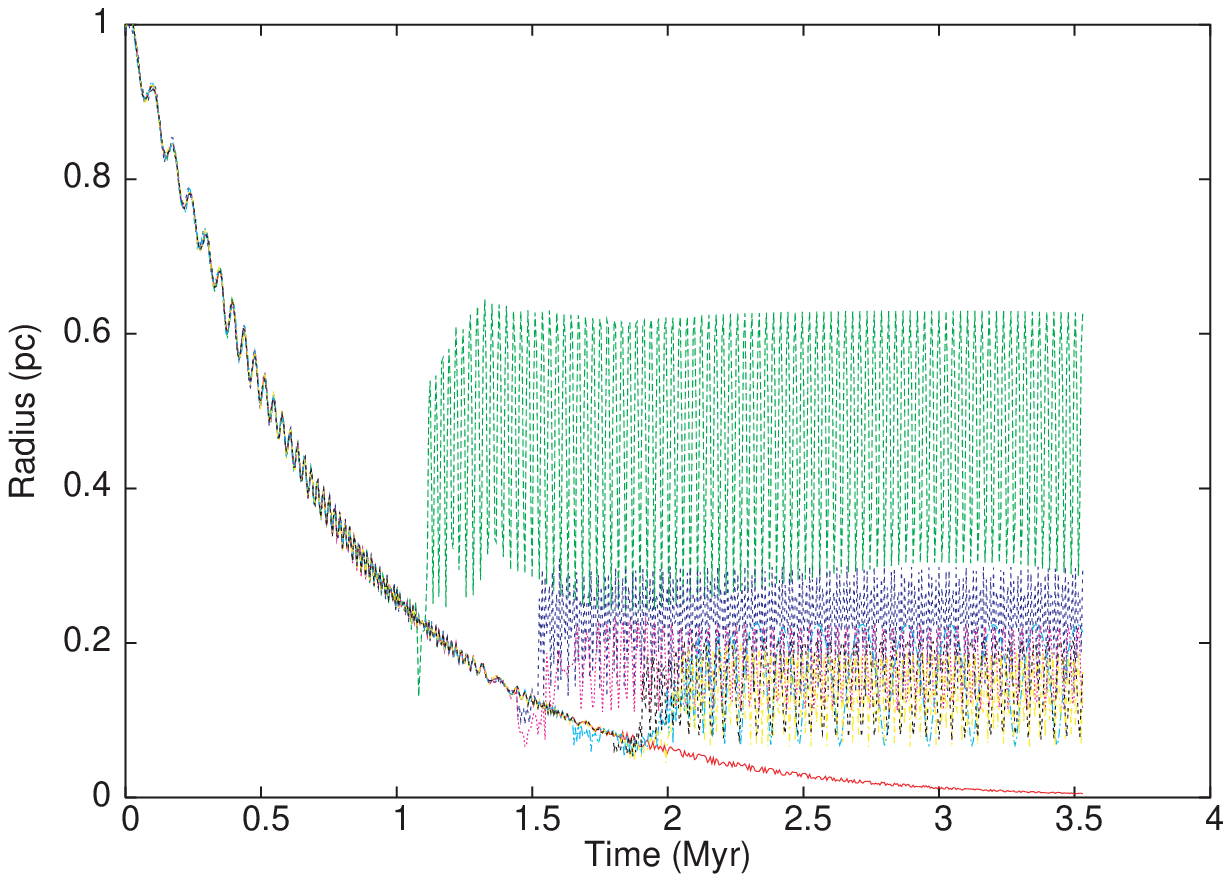}}
\end{tabular}
\end{center}
\figcaption{
Transport of multiple stars to the inner few tenths of parsecs by the IMBH.  The left panel shows the transport in a simulation that includes a Bahcall-Wolf $r^{-1.75}$ cusp about the SMBH, while the left panel lacks such a cusp.  Stripped stars will initially follow the IMBH in its orbit by being resonantly trapped at the IMBH's $L_5$ point.  The pericenter distance soon after stripping is roughly the semimajor axis.  A subsequent IMBH encounter will scatter the star into a more eccentric orbit, with the semimajor axis increasing as well.  In both simulations the IMBH mass is $4105 M_{\odot}$ and initial eccentricity $0.25$.
\label{fig:trans0}
}
\end{figure*}

The presence of an IMBH is a sufficient mechanism for transporting stars deep into the potential well of the SMBH.  In most cases, the infall of the cluster core causes several stars ($10-20\%$ of the original core population) to be transported within $0.3\pc$ of the Galactic Center, and in some cases, further.  Figure~\ref{fig:trans0} shows a typical set of stars which are carried into the potential well.  Some stars are transported into small semimajor axis orbits but then are perturbed into larger orbits due to IMBH scattering.  Due to the strong tidal field, the IMBH typically loses most of its stars by $0.1\pc$, and falls in alone.  

In principle, simple analytical arguments based on tidal radii would allow transport interior to this radius, for more tightly bound systems.  However, strong internal dynamics in the core coupled with the inspiral of the IMBH will regulate this effect.  The dynamics promote an overall expansion of the system, so that tightly bound core remnants will tend to have their haloes removed by the tidal field.  The subsystem then binds further, stengthening the dynamical interactions, possibly leading to the perturbation of a member into a halo orbit.  This member is pared from the system by the tidal field, which becomes stronger due to the IMBH inspiral, and the core is bound tighter.  This process continues until there is only one star bound to the IMBH.  The consequence of this process is that the delivery of multiple stars with small semimajor axes can be damped, since the strong perturbations which kick stars out of the core will push them into larger orbits.  Therefore, the deposition of the S-stars by a single IMBH passage seems improbable within this scenario. 

\begin{figure}
\plotone{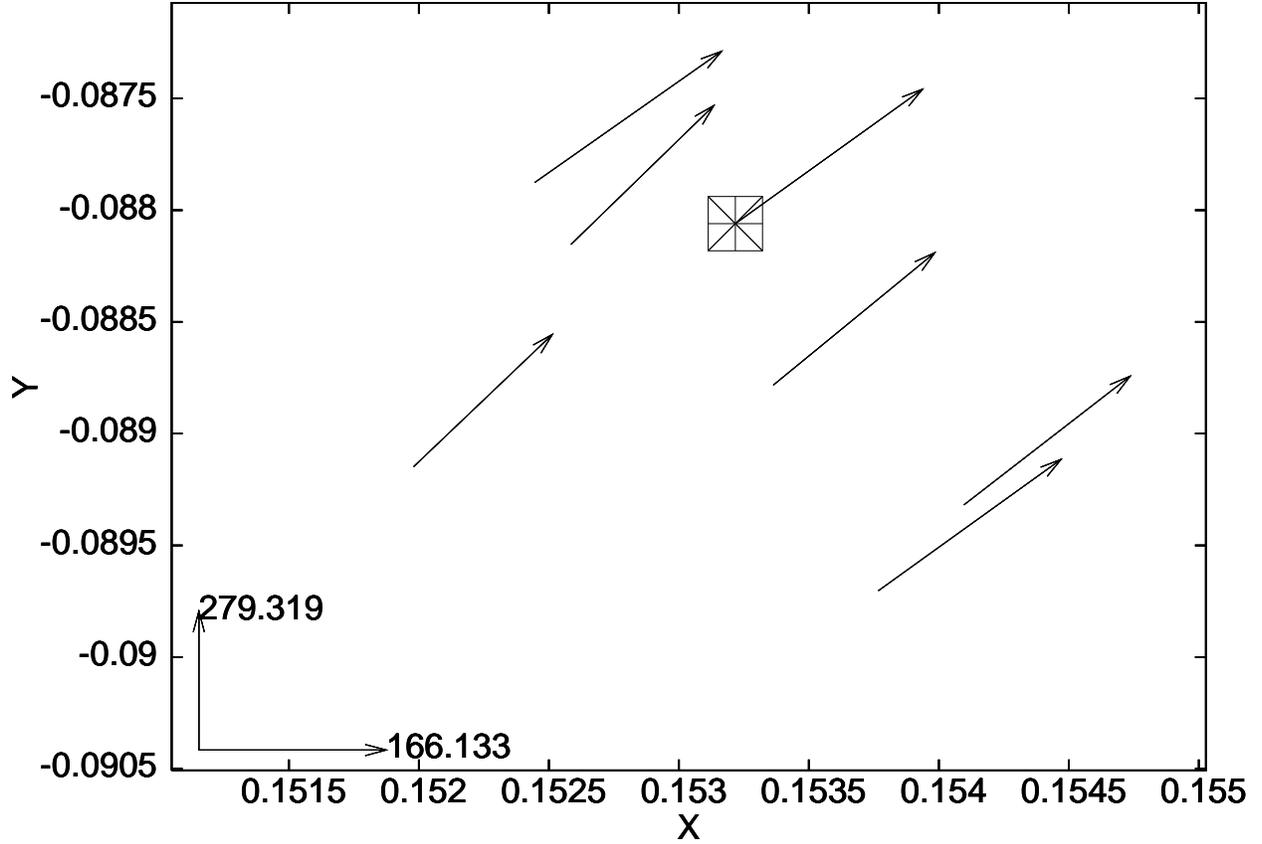}
\figcaption{
A typical simulated comoving group.  The arrows in the lower left corner indicate the x- and y-velocities of the stars.  The position of the IMBH is indicated by the box.  Note that the x- and y-positions are relative to the origin, where the SMBH sits;  this comoving group sits about $0.17\pc$, or $4''$, from the GC.  The diameter of this group is about $10^{-3}\pc$, while the Roche envelope, with radius $0.018\pc$, lies far outside the boundaries of the plot.  This typical structure resembles the IRS13E complex reported by Paumard et al. (2004).
\label{fig:com}}
\end{figure}

Before all the stars get stripped, there will be a period in which a small ($N\sim 5$) number of stars remain bound to the IMBH.  These configurations could be considered ``comoving'', since while these stars orbit the IMBH, the IMBH orbital velocity dominates that of the stars, and an observer looking at such a configuration from afar would see small variations in the stellar velocities when viewed as a group.  An example of this is seen in Figure~\ref{fig:com}, for the case in which there is no Bahcall-Wolf cusp or mass distribution.  

The important parameter for the endurance of comoving groups is obviously the IMBH mass, because of the implied larger Roche lobe.  There is no evidence of any effect of the cusp, but that is because the cusp is a continuous structure seen by the cluster core, instead of a more realistic set of disretized masses.  In Figure \ref{fig:trans0}, the IMBH transports stars into orbits of semimajor axis of roughly $0.2\pc$, but even manages to retain multiple stars further in, whether there is a cusp (left panel) or not (right panel).  

Observationally, one would expect that the IMBH is unlikely to be detected directly, but its presence inferred from the gravitational hold it retains on its orbiting stars.  In Figure~\ref{fig:com}, we see a typical comoving group from above the inspiral plane.  Astrophysical units are used, and the box denotes the IMBH.  The radius of this structure is about $\sim 10^{-3}\pc$, is located $0.17\pc$ from the GC, and is being held together by an IMBH of mass $4062 M_{\odot}$.  The edge of the Roche envelope of the IMBH lies $0.018\pc$ from the IMBH, so these stars lie deep within the Roche lobe.  Further, analysis of the stellar orbital energies shows that these stars are gravitationally bound to the IMBH.  Such structures form with some frequency, in this scenario. 

Table~\ref{tbl:com} shows how the different simulation parameters affected the efficiency of transport.  Note that for the modelled masses, the minimal semimajor axis ($0.129\pc$) to which $6$ or more stars are transported corresponds to a IMBH Roche lobe of $0.0145\pc$.  Two trends are in evidence: only IMBHs more massive than about $4000 M_{\odot}$ can transport multiple stars within $\sim 0.13\pc$, and while massive IMBHs can make such deliveries, they must do so when their eccentricities are moderate, as no simulation with $e=0.5$ transported stars deeper than $0.185\pc$.  These two points may constrain the applicability of the cluster scenario to the Galactic center.

Note that the IRS16SW comoving group has been recently identified less than two arcseconds ($< 0.08 \pc$) from Sgr A*, while the IRS13E comoving group is about four arcseconds ($\sim 0.16\pc$) away.  The current simulations are able to explicitly replicate structures similar to the latter, but not the former, due to weaknesses in the numerical treatment.  However, it is unclear which portion of the IRS16SW orbit has been observed; it may be near periastron on an eccentric orbit, in which case its semi-major axis could be as much as a factor of two larger.  While these simulations don't reproduce IRS16SW, they suggest that it could be formed given more tightly bound initial cores, which are difficult to simulate with current numerical techniques.  

\subsubsection{Disks}
\label{sec:disks}
One goal of this effort is to understand if repeated few-body encounters in strong tidal environments can produce multiple disk populations, possibly counterrotating and/or inclined relative to one another, on realistic timescales.  In the Galactic Center, two such disks are claimed to exist inside of $0.4\pc$, and their existence is not understood.  

\begin{figure}
\plotone{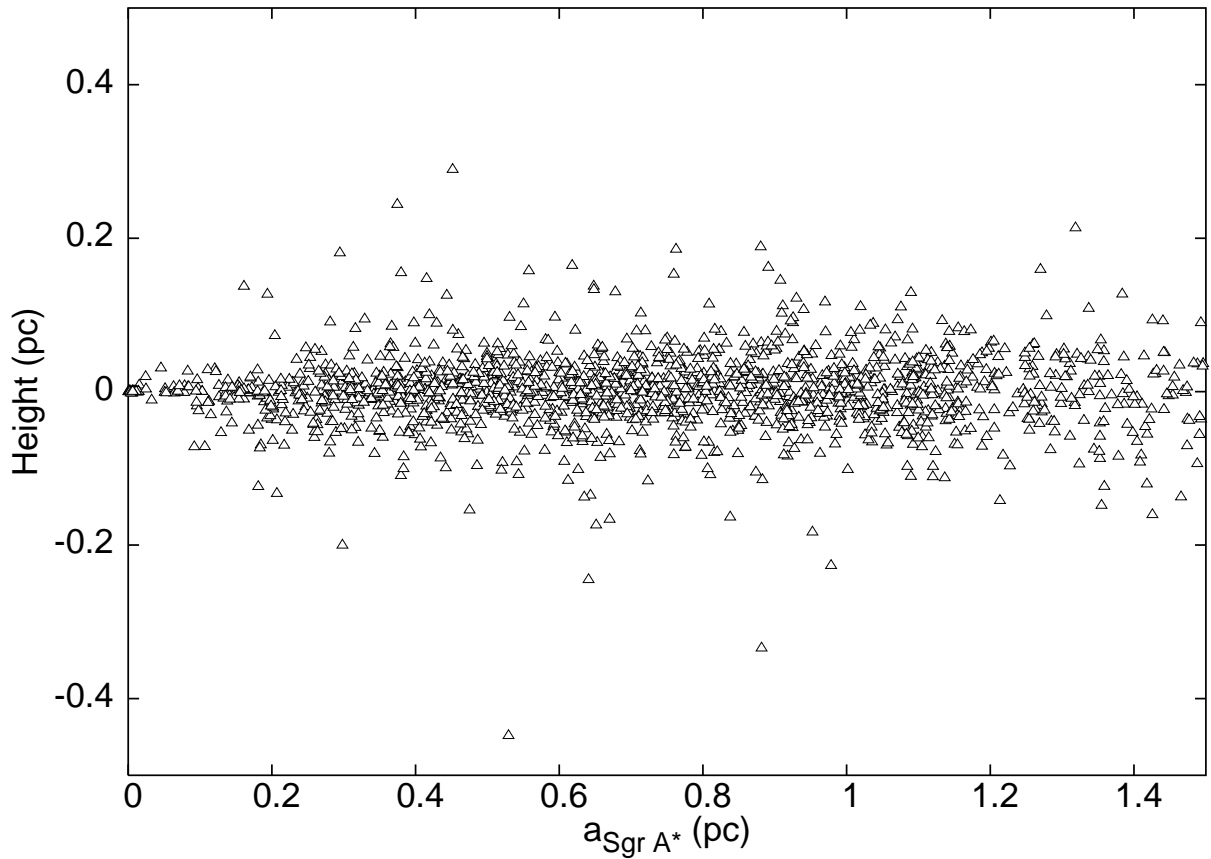}
\figcaption{Radial profile of all disks formed during simulations including the realistic cusp.  No difference was seen in the no cusp case, and so it is not plotted here.  This is an aggregate of all data for these simulations.  The mean opening angle for the disk stars is $\sim 13\dgrs$, regardless of the initial phase space coordinates, IMBH eccentricity or mass.  
\label{fig:disk1}}
\end{figure}

In principle, as the cluster core spirals in, tidal stripping produces a tail or wake of stars.  The IMBH orbit dominates the energy and angular momentum of the stellar orbits, and so when stars are stripped they are strewn initially into orbits that reflect the environment in which they were stripped, i.e., the inclination, semimajor axis, and eccentricity of the IMBH.  As a consequence, neglecting subsequent encounters, the disks that are produced should be vertically thin.  Figure~\ref{fig:disk1} shows the side profile ($R-z$) of all stripped stars resident in disks.  Note that post-stripping encounters can significantly perturb the stellar orbits and erase this history of inspiral.  Typical star-star interactions are too weak to produce significant variations in orbital parameters, even considering the large number of weak impulses that occur over the long random walk these parameters endure during their lifetime in the disk. Thus, only IMBH-star interactions are important in randomizing the distribution of stars in the disk.

\begin{figure}
\plotone{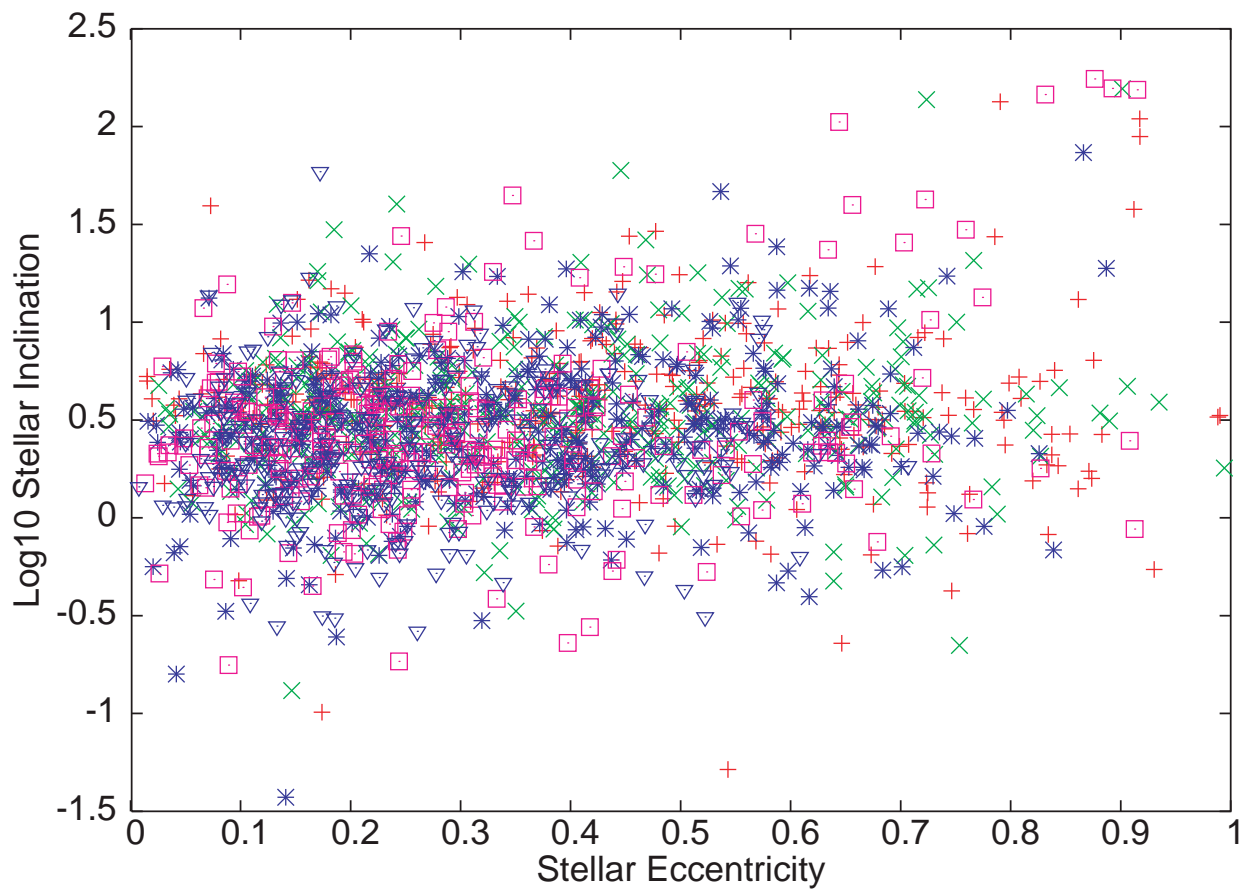}
\figcaption{The effect of IMBH mass on final stellar inclination and eccentricity.  This is an aggregate of all data from the simulations; the different colors reflect different IMBH masses in the range $1015-5540 M_{\odot}$.  Little variation due to mass is seen, indicating that the number of strong star-IMBH interactions is few.\label{fig:iemass}}
\end{figure}

The strength and direction of these impulses determines the effect of the perturbation.  Because the disks should be thin, there are typically no large normal forces involved in an interaction since the stars and IMBH are in nearly the same plane.  Thus, one cannot expect that many large inclination orbits will be produced, even by strong interactions.  Indeed, Figure~\ref{fig:iemass} shows that there is no difference in the final eccentricities and inclinations of stars, regardless of the IMBH mass used.  If strong perturbations due to IMBH were occurring frequently, there would be some scaling of these values with that mass.  Furthermore, resonant trapping by a migrating body can, in principle, create counter-rotating disks (Yu \& Tremaine 2001).  While some particles are trapped temporarily in our simulations, this mechanism does not appear to operate with any significant efficiency.

\begin{figure}
\plotone{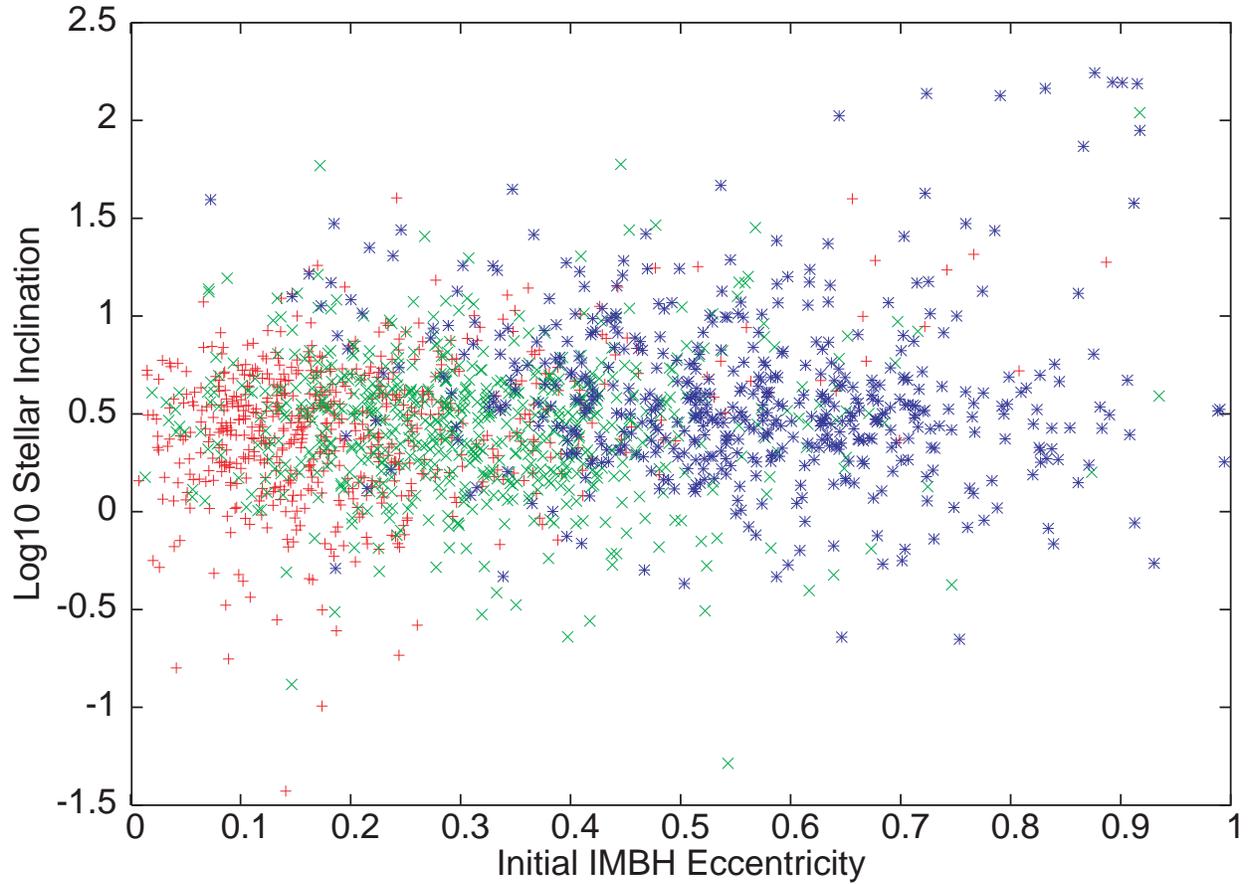}
\figcaption{ Stellar inclinations versus initial IMBH eccentricity.  The initial IMBH eccentricity is not an important parameter in causing strong star-IMBH scatterings normal to the IMBH orbital plane.  Red plusses correspond to IMBH $e=0$, green crosses, $e=0.3$, and blue stars, $e=0.5$.
\label{fig:eeoi}}
\end{figure}

Since the disks are thin, the formation of two distinct, counter-rotating, inclined disks is unlikely.  However, perturbations do occur, and stars undergoing strong interactions typically end up on high eccentricity, high inclination orbits.  In Figure~\ref{fig:eeoi}, the upper right corner contains several stars (order $10$ out of $\sim 2000$) that have such orbits.  This ratio (order $0.5\%$) is what one would expect from simple analytical estimates.

\begin{figure}
\plotone{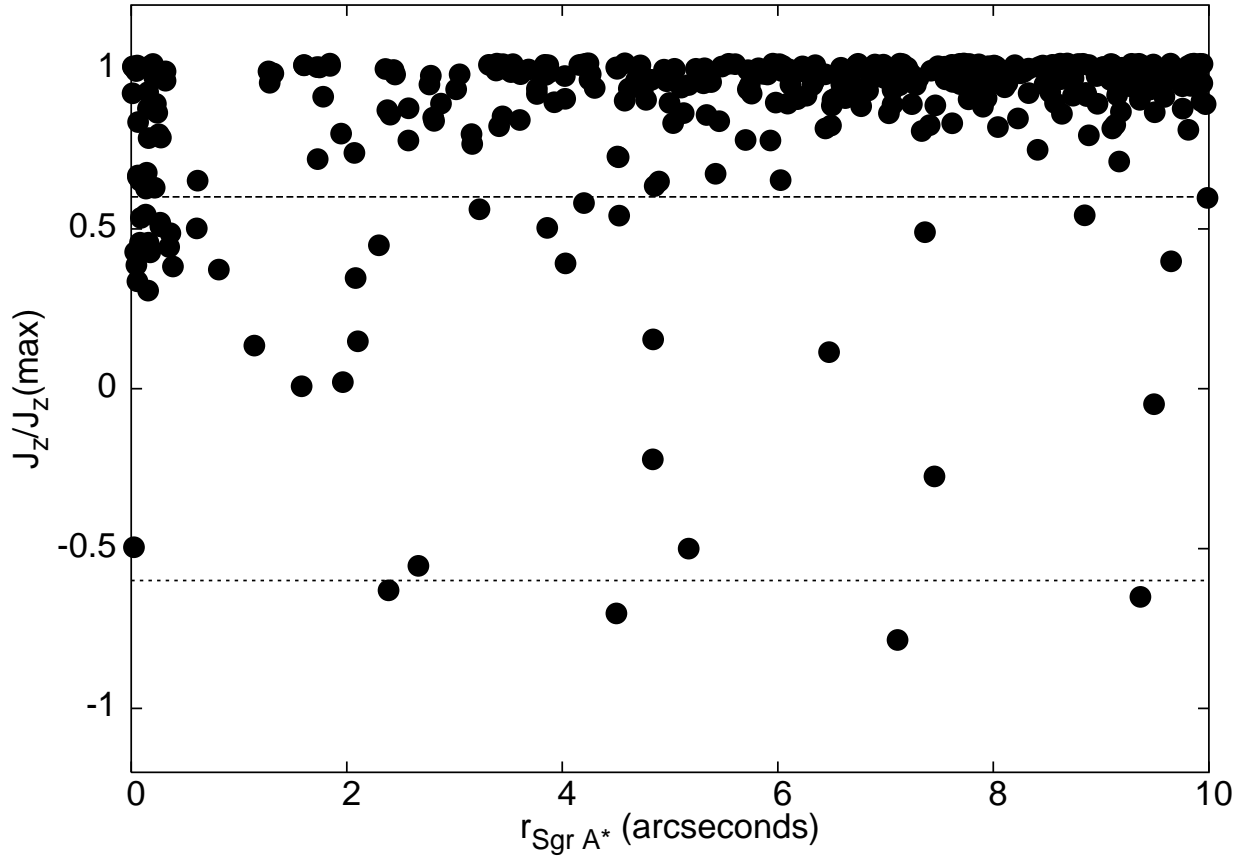}
\figcaption{Normalized angular momentum versus distance to the SMBH, for all stars in the no-cusp simulations.  Stars moving clockwise comprise nearly $98\%$ of stars in the sample, and the mean disk opening angle is $\sim 13\dgrs$.  The stars furthest to the left are those transported deep into the GC as singletons orbiting the SMBH, and are not included in the statistics.  Note the significant depletion of particles between $0.5''$ and $1.5''$, similar to the inner edge reported by Paumard et al. (2006).
\label{fig:jz}}
\end{figure}

While no multiple disk systems were produced, many of the parameters of the typical remnant disk are consistent with the individual observed disks.  Figure~\ref{fig:jz} shows the normalized angular momentum 
\begin{equation}
\frac{J_z}{J_z(max)} = \frac{xv_y-yv_x}{rv_r}
\end{equation}
against the radial distance from the origin, out to $0.4\pc$.  This analysis is similar to that of Genzel et al. (2003).  Most stars in the sample are rotating in the same direction with only $1\%$ of stars on counter-rotating orbits.  The mean disk opening angle for simulations with no cusp is $13.8\dgrs \pm 2.9\dgrs$, and for simulations including the cusp, the mean disk opening angle is found to be $12.6\dgrs \pm 2.8\dgrs$, so there is no statistical difference between the two.  In the figure, the points farthest to the left are singleton stars transported deep into the potential well by the IMBH; these were not included in the means calculated above.  Note that there is a gap near $0.05\pc$, which could correspond to the possible inner edge reported by Paumard et al. (2006).  This may reflect the formation of a standard core-halo separation in the cluster, formed as a result of internal dynamical relaxation.  The inner edge of the disk would then correspond to the removal of the last of the more loosely bound halo stars, which follows naturally from the self-regulation mechanism addressed in \S~\ref{sec:trans}.  More experiments are needed to understand this in greater detail.

\begin{figure}
\plotone{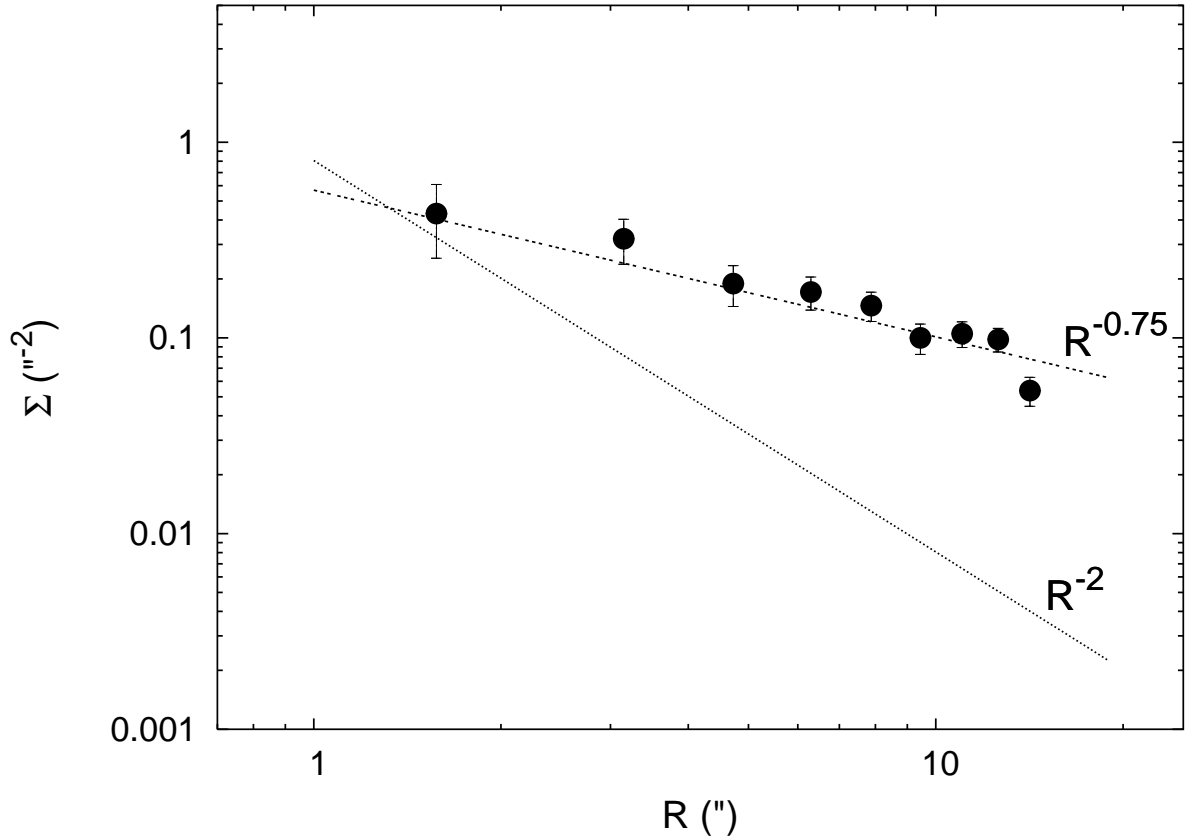}
\figcaption{Comparison of simulated disk surface density profiles with observation.  Regardless of the initial profile used, the stripped stars relax into $\Sigma\propto r^{-0.75}$ profiles.  This starkly contrasts with the observed $\Sigma\propto r^{-2}$ profile reported in Paumard et al. 2006.  The error bars are based on Poisson statistics, and do not reflect some systematic uncertainty.
\label{fig:sig.eps}}
\end{figure}

The cluster simulations produce remnant stellar disks with surface density $\Sigma \propto r^{-0.75}$.  This result appears to be robust with respect to the initial density profile used.  Figure~\ref{fig:sig.eps} shows a comparison of the surface density obtained from the simulations to that of Paumard et al. (2006).  A linear fit of the simulated surface density profile finds a best-fit power-law slope of $\alpha=0.74\pm 0.05$.  We also tested three other profiles: $1/r$, $1/r^3$, and a Plummer sphere, all three producing final profiles similar to the $r^{-0.75}$.  Note that internal dynamical evolution will drive the density profile towards the Bahcall-Wolf law (Bahcall \& Wolf 1976; Baumgardt, Makino, \& Ebisuzaki 2004).  This, followed by tidal stripping via a Roche criterion and deposition into a disk would imply a density $\Sigma \propto r^{-0.75}$.    

Our finding contrasts with the observed profile of Paumard et al. (2006), who find $\Sigma \propto r^{-2}$. These observations are limited in several respects; most notably, the magnitudes are limited to about $K=13.5$, which therefore includes only the most massive stars, and full three-dimensional orbit information is incomplete with large uncertainties.  It is not obvious that any of these limitations are responsible for the disagreement in surface density profile.  However, if this discrepancy continues to hold as more refined observations are made, the comparison with our simulated surface density profiles will be a significant constraint on the cluster infall scenario.

As mentioned previously, the IMBH is responsible for assigning orbital parameters to stars immediately after they are stripped, subject to a small scatter about the mean represented by the IMBH values.  Their resulting orbits therefore represent a fuzzy memory of the IMBH infall, providing a constraint on the history of the GC.  In particular, this shows that the two, roughly coeval, counter-rotating disks at the Galactic center cannot have been formed through the inspiral of one cluster.

\subsubsection{Scattering during inspiral}
\label{sec:scatter}
\begin{figure*}
\begin{center}
\begin{tabular}{cc}
\resizebox{80mm}{!}{\plotone{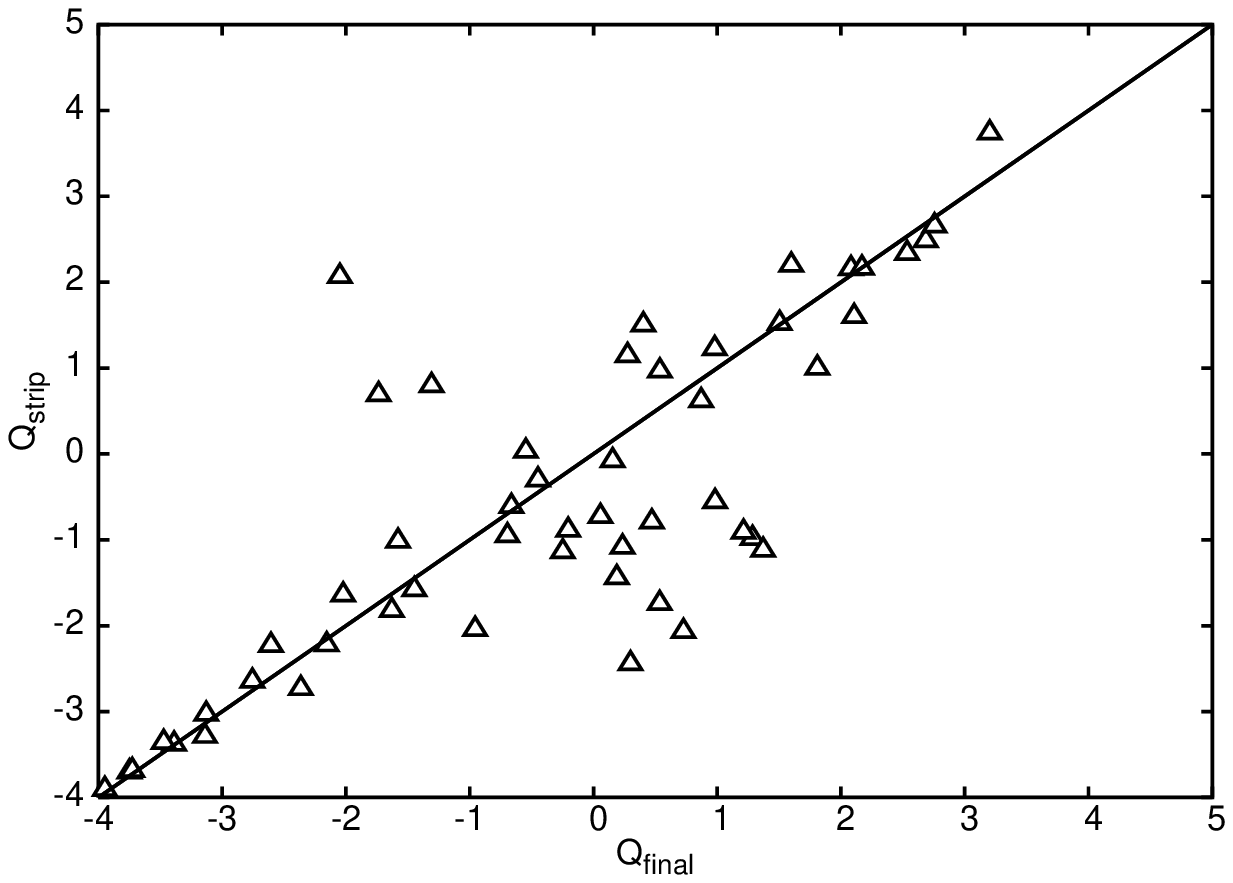}} &
\resizebox{80mm}{!}{\plotone{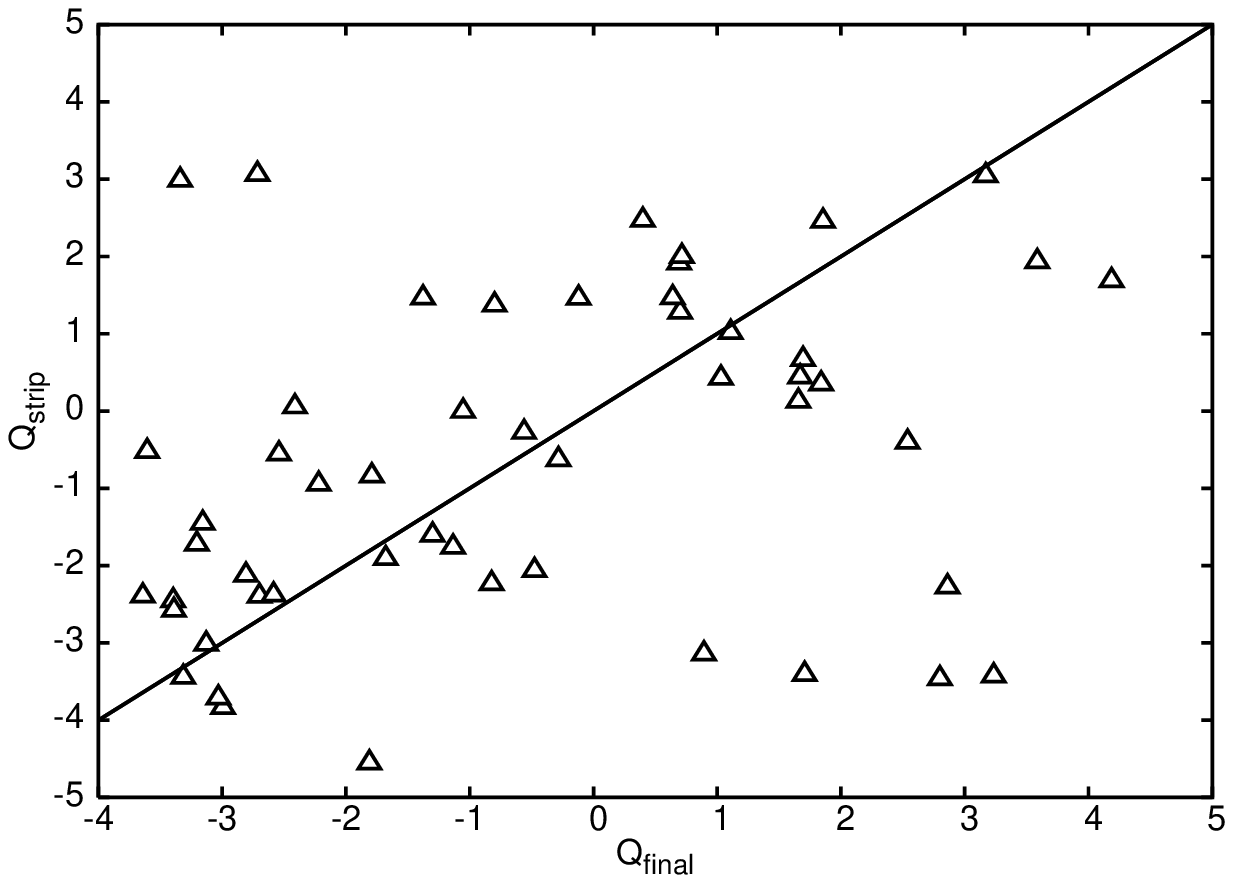}}
\end{tabular}
\figcaption{Comparison of the values of the Tisserand relation $Q=1/2a+\cos{i}\sqrt{a(1-e^2)}$ immediately after stripping and at the end of the simulation.  Left panel: IMBH mass $4062 M_{\odot}$, $e=0$.  Right panel: IMBH mass $1357 M_{\odot}$, $e=0$.  The higher mass case has less scatter because, although a higher mass would cause stronger interactions than a lower mass, $\dot{r}$ is larger, and so the IMBH will not have as much opportunity to interact with the disk stars.  Lower mass IMBH do not descend the potential well as quickly, allowing the possibility for more interactions.
\label{fig:sca}}
\end{center}
\end{figure*}

The results of these simulations have already provided insight into the nature of the environment in which stripped stars may live as the IMBH continues its plunge.  Multiple lines of evidence suggest that weak scattering is common once stars are removed from IMBH orbit.  On a global scale, Figure~\ref{fig:sca} shows the relationship between the Tisserand relation immediately after stripping and at the end of the simulation.  The Tisserand relation $Q=1/2a+\cos{i}\sqrt{a(1-e^2)}$ is an approximate expression of the Jacobi constant in terms of the orbital parameters, and its variation reflects the effect of orbital perturbations due to a passing encounter.  (e.g. Murray \& Dermott 1999)  Note that there are three regimes, in the left panel.  The bottom left is composed of stars with high inclinations, the top right is of stars with small semimajor axes, and the middle of the plot contains the rest.  Clearly, this middle section is a 'hot-zone' for perturbations, since the other two regimes are outside the domain of strong IMBH influence.  Strong scattering events clearly occur and change the value of the orbital elements.  Evidence has been presented indicating that higher mass IMBHs are more efficient at transporting stars, but here, it is seen that they are less likely to provide significant perturbation to the stars, while lower mass IMBH may do a better job.  In the context of the Galactic center, IMBHs are a pertinent mechanism for comoving groups, but are unlikely to explain the randomization of orbits seen there. 

\begin{figure*}
\begin{center}
\begin{tabular}{cc}
\resizebox{80mm}{!}{\plotone{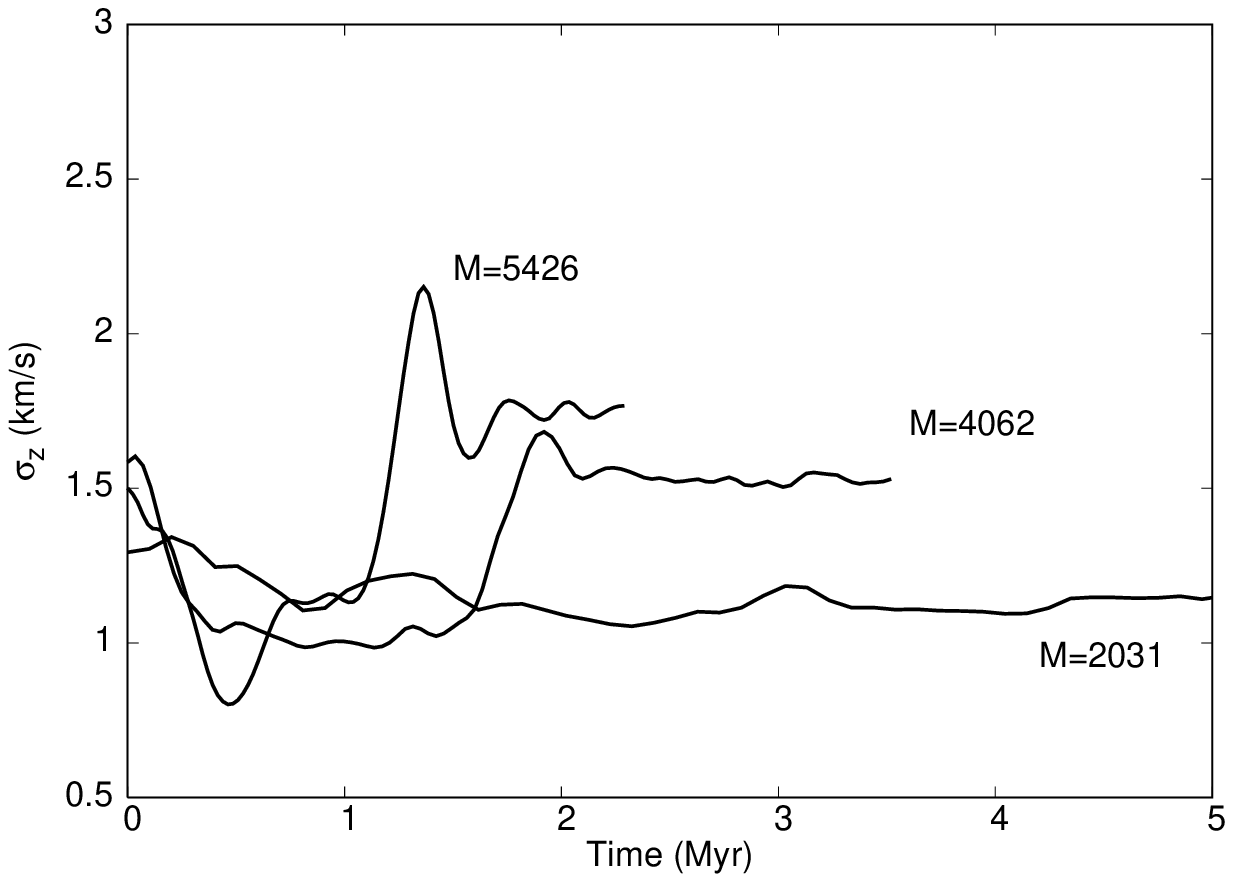}} &
\resizebox{80mm}{!}{\plotone{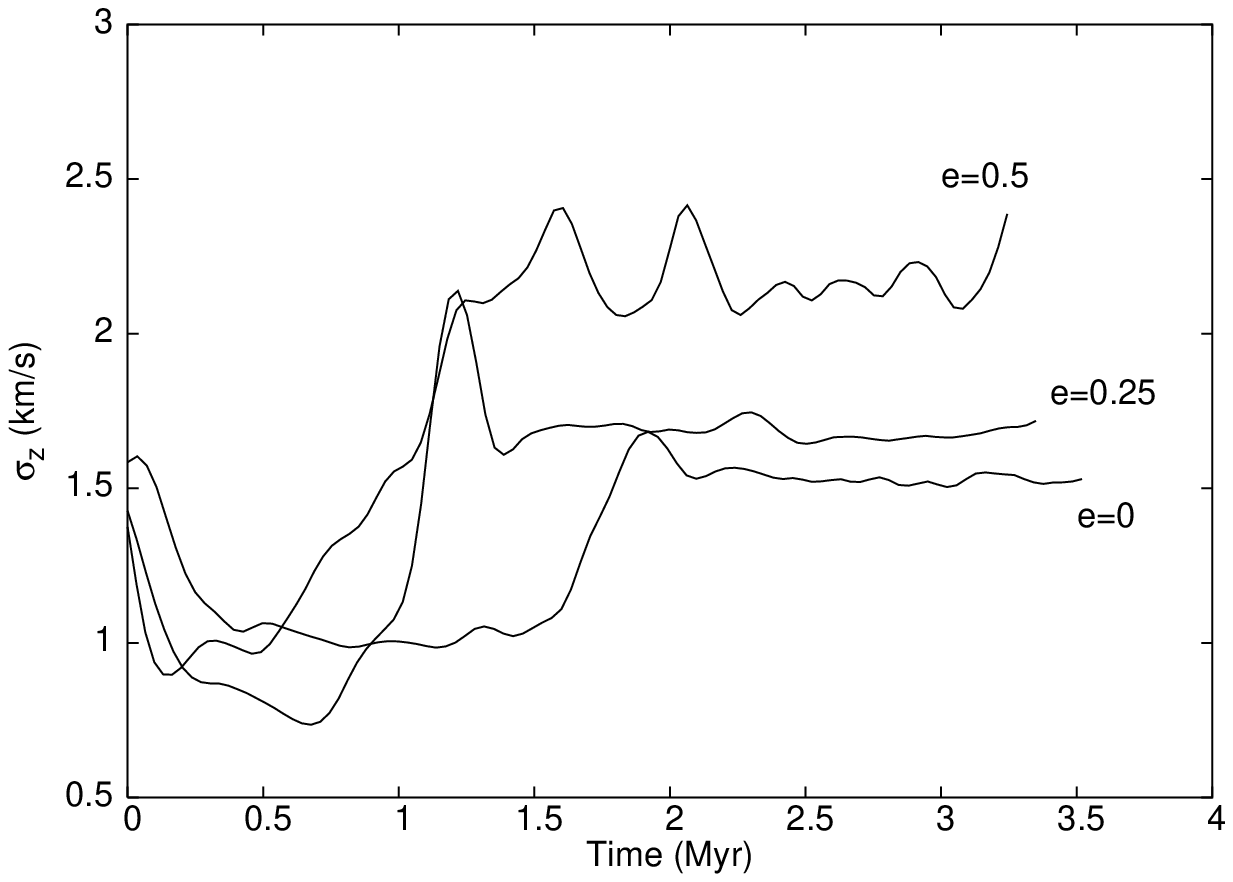}}
\end{tabular}
\figcaption{
Trends in the evolution of $\sigma_z$, the z-component velocity dispersion, of stripped stars.  Since the dominant motion is in the $x-y$ plane, $\sigma_z$ provides the best estimate of the change in dispersion over time.  Parameters for different runs are labeled, and the curves have been smoothed in order to view the basic trends.  Left panel: mass varied with eccentricity $e=0$.  Right panel: mass constant ($4062 M_{\odot}$) and eccentricity varied.  
\label{fig:veldisp}}
\end{center}
\end{figure*}

\begin{figure*}
\begin{center}
\begin{tabular}{cc}
\resizebox{80mm}{!}{\plotone{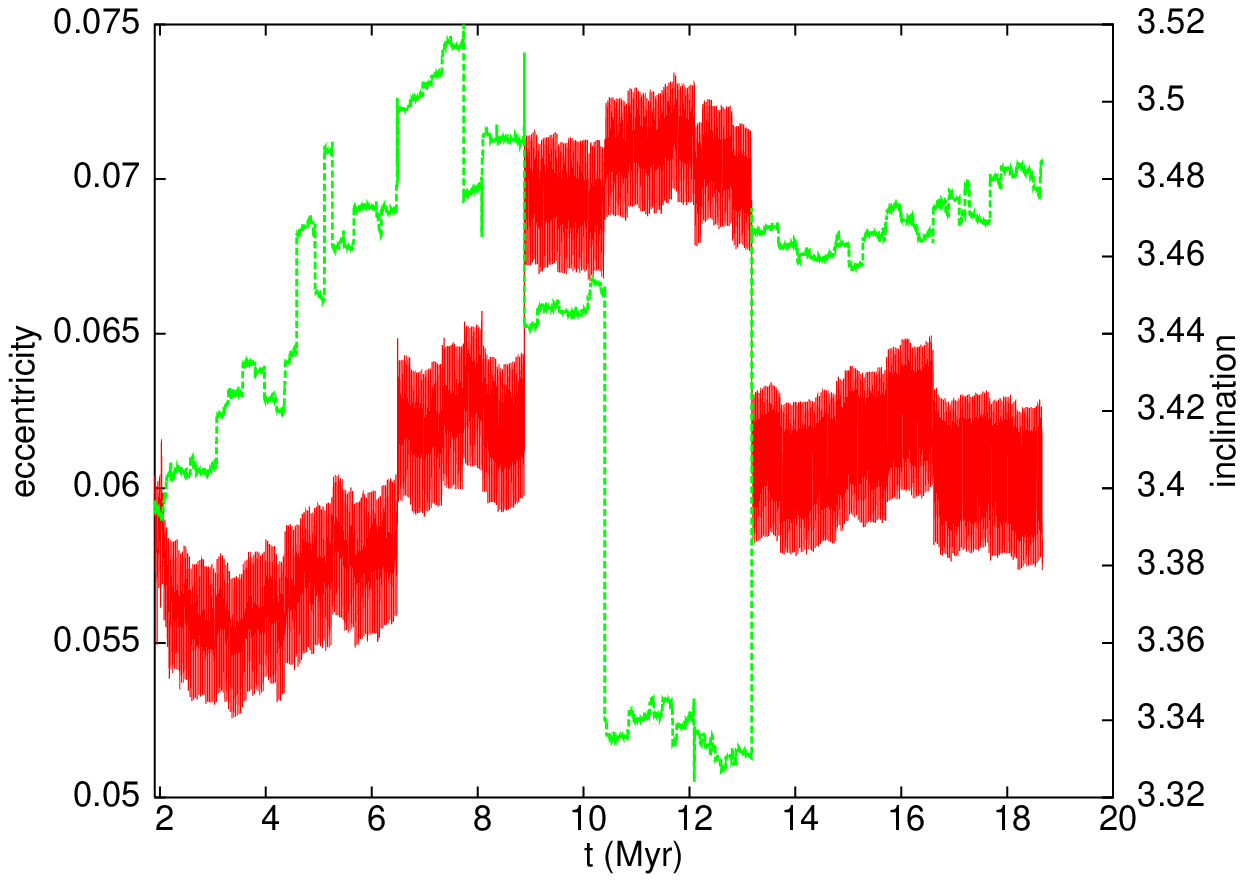}} &
\resizebox{80mm}{!}{\plotone{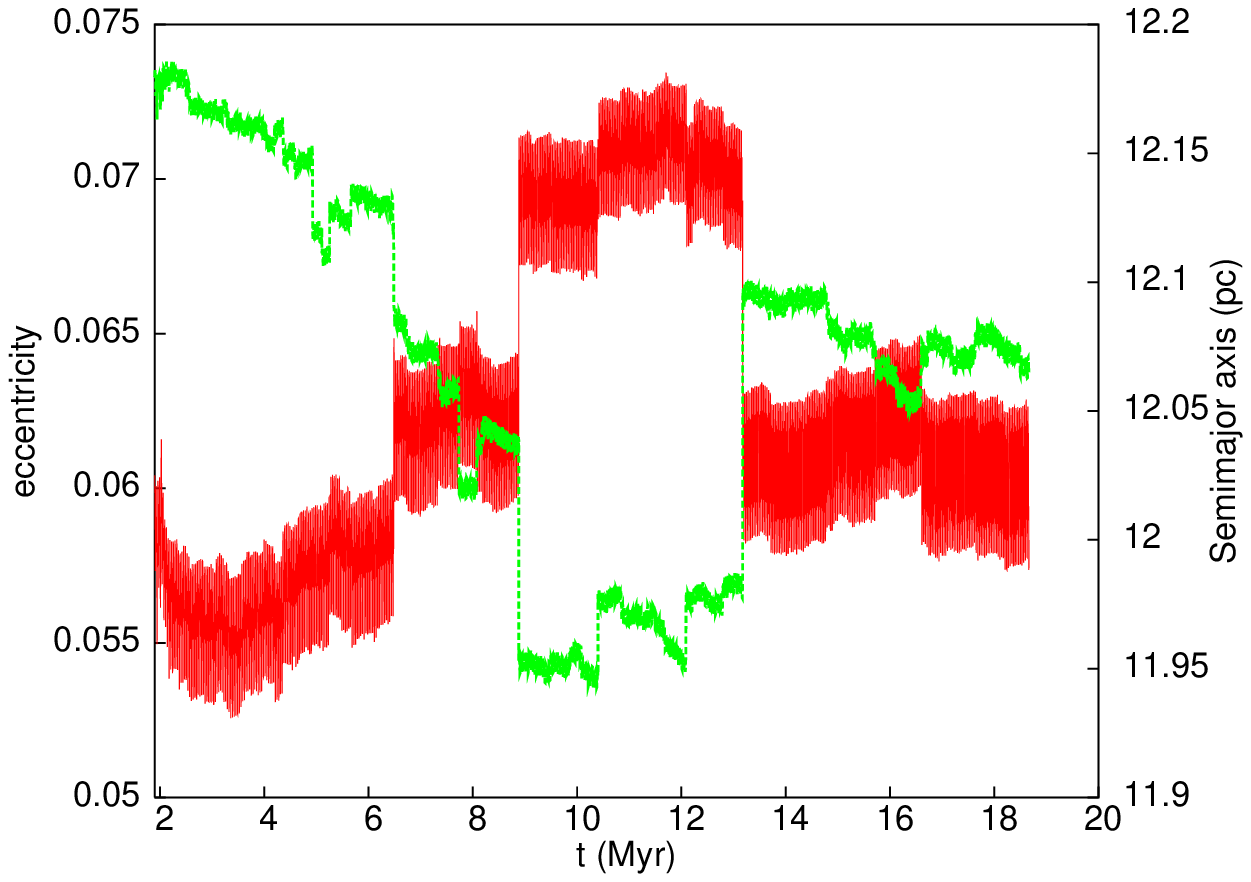}}
\end{tabular}
\figcaption{The effects of interactions on the orbital elements.  The thick band in both plots is the eccentricity.  The thin jagged lines in each plot are smoothed for clarity; (left panel) inclination, and (right panel) semimajor axis.  The star undergoes numerous weak encounters that perturb its orbit slightly, sending the orbital elements into a random walk.  Here, the star suffers a number of weak encounters that drive eccentricity up, then two successive stronger encounters that drop the value down.  At the same time, the stronger encounters serve to dramatically decrease, then increase, the inclination and semimajor axis.
\label{fig:scatter}}
\end{center}
\end{figure*}

Figure~\ref{fig:veldisp} shows the evolution of the z-component velocity dispersion $\sigma_z$ for stripped stars for a typical set of simulations, covering a range of eccentricities and inspiral times.  In the left panel, runs with different IMBH masses but zero eccentricity are compared, and in the right panel, the IMBH eccentricity is varied holding the mass constant ($M=4062 M_{\odot}$).  The initial large increases are due to the loss of stars from the cluster, and addition to the list of stripped stars.  After most or all stars are stripped, $\sigma_z$ remains fairly flat.  The large spikes occur when tightly bound stars are stripped or interact through few-body interactions to leave the cluster at high velocity.  This is typically followed by the loss of most of the remaining core members, and $\sigma_z$ flattens.
On a smaller scale, an analogous and revealing set of evidence comes from an analysis of the changes in the orbital elements of typical stars.  Most stars will experience $\sim 20$ strong scatters per $10 \Myr$ simulation, and the nature of these interactions is reflected in how $a$, $e$, and $i$ change.  Figure~\ref{fig:scatter} shows the effect of interactions on a typical orbit.  Numerous weak encounters force the orbital parameters into random walks, with stronger encounters more dramatically changing the values.  As with any random walk, the final distribution of orbital parameters is uncertain, and is dependent on the nature of the perturbations.

\begin{figure}
\plotone{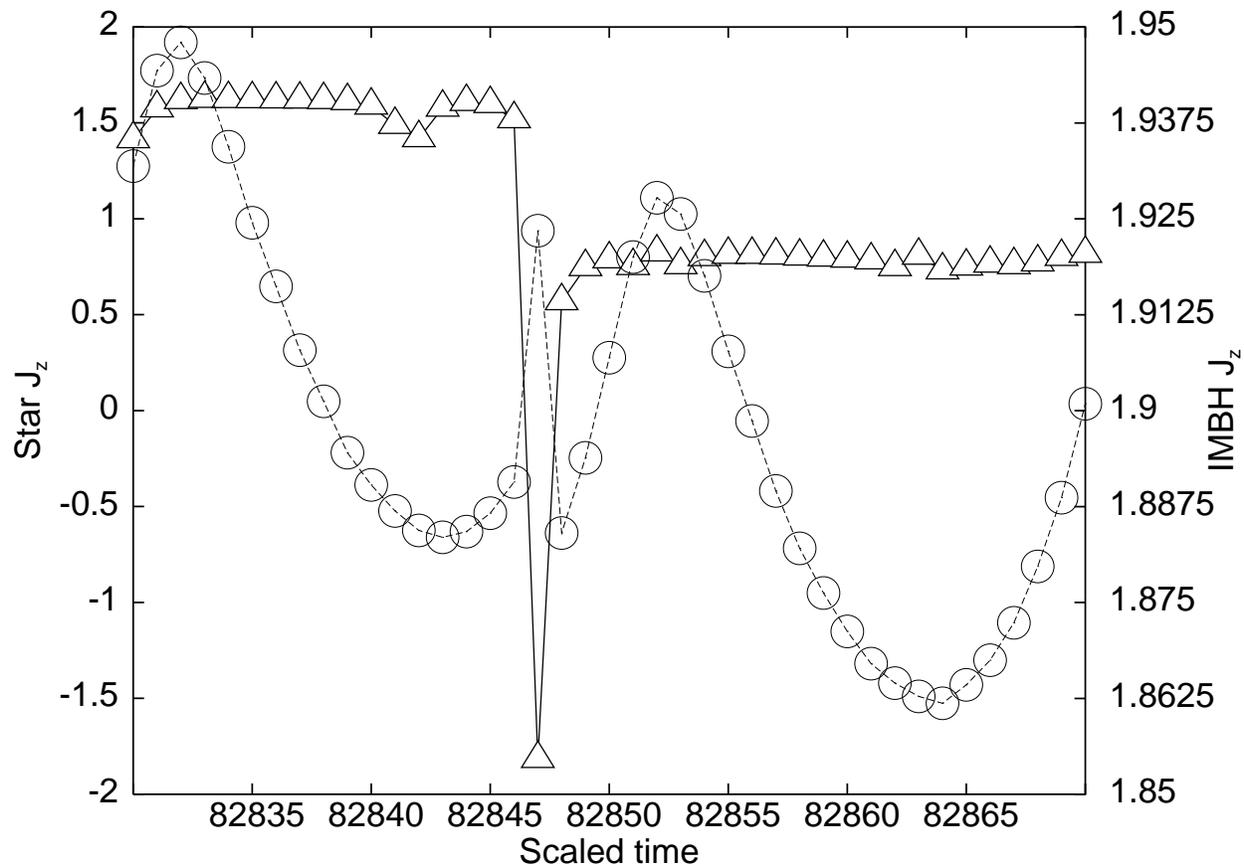}
\figcaption{
The evolution of the z-component of angular momentum for a star (L) and the IMBH (R) during a near passage.  The bottom axis is in scaled, N-body time units.  Note the exchange of $J_z$ during the encounter and the subsequent recovery.  The IMBH orbit was not strongly affected by this; the magnitude change is only about $2\%$, but the star is strongly affected, with a $200\%$ change during the spike, and final $62.5\%$ change after the encounter. 
\label{fig:7pert}}
\end{figure}

\begin{figure*}
  \begin{center}
    \begin{tabular}{cc}
      \resizebox{70mm}{!}{\includegraphics{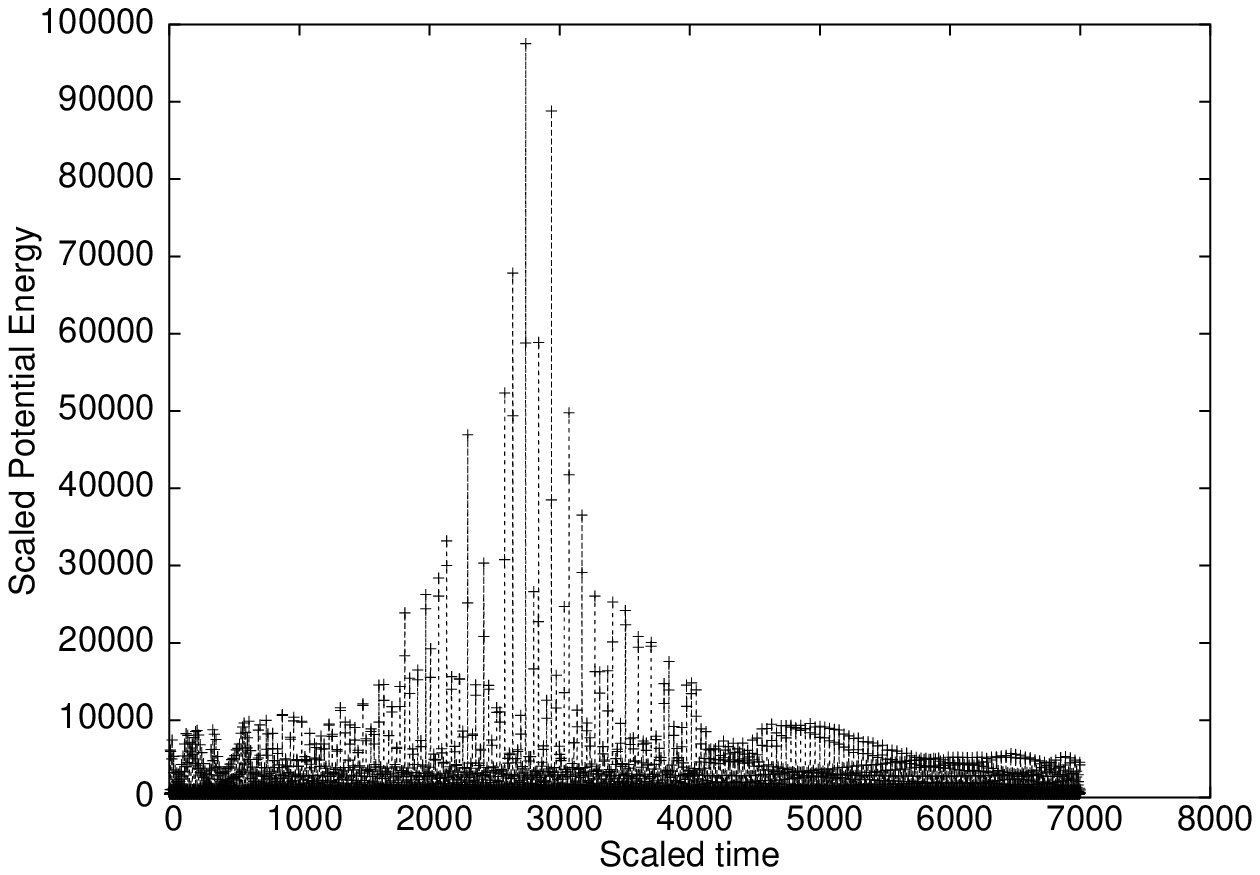}} &
      \resizebox{70mm}{!}{\includegraphics{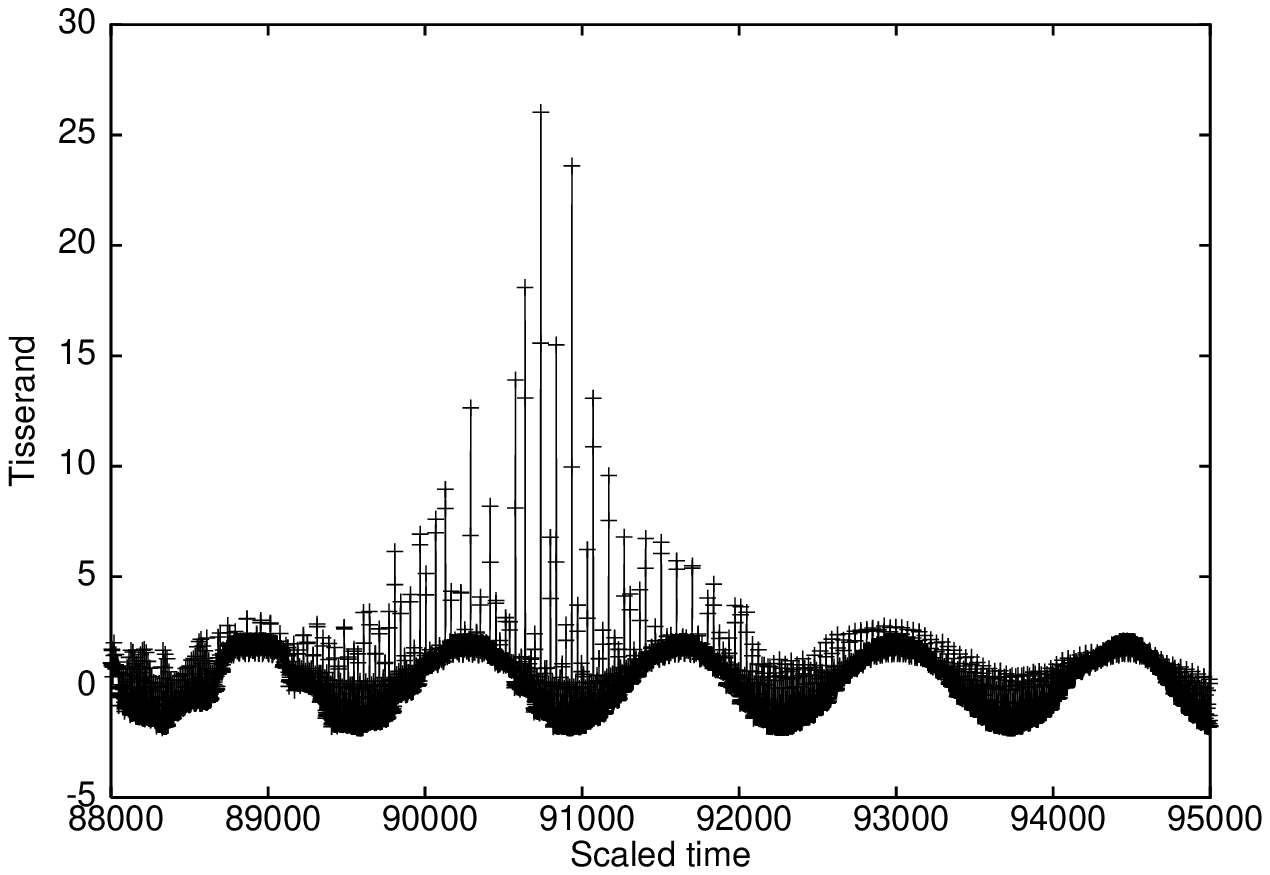}}
    \end{tabular}
 \figcaption{
Effect of subsequent strong IMBH encounter on star orbit.  Left panel shows the star potential energy, with several near-order-of-magnitude spikes evident.  Right panel shows the Tisserand relation $1/2a+\cos{i}\sqrt{a(1-e^2)}$, clearly showing strong changes in the orbital parameters during this time.  This star ends with a stable counter-rotating orbit with inclination of $115\dgrs$.
\label{fig:morepert}}
\end{center}
\end{figure*}

Strong scattering is the cause of the counter-rotation of stars seen in several simulations.  The culprit is usually a near passage of the IMBH.  Figure~\ref{fig:7pert} shows the effect of the passage of one of the stars near the binary containing the IMBH.  The star nears the binary and is subject to a strong torque.  This perturbation is strong enough to send the star into a complex orbit, in which it is subject to additional strong perturbations and passages near the SMBH, which apply additional torque and result in a large inclination orbit.  Figure~\ref{fig:morepert} shows a short time period after the IMBH encounter, when the star has been sent into a difficult orbit, with changing peribothron.  Its potential energy spikes, and a plot of the Tisserand $Q$ shows the net orbital changes that occur.  At the end of this simulation, this star attains an orbital inclination of $115\dgrs$.  This process is responsible for the orbital configurations of most of the counter-rotating stars.  It needs to be emphasized that such events are rare; however, this process will tend to randomize the orbits of stars near the Galactic Center.  In the present simulations, approximately $1\%$ of stars find themselves in retrograde orbits at the end of the simulation, and in no circumstance are ``counter-rotating disks'' created.  

\subsection{Stripping}
\begin{figure}
\plotone{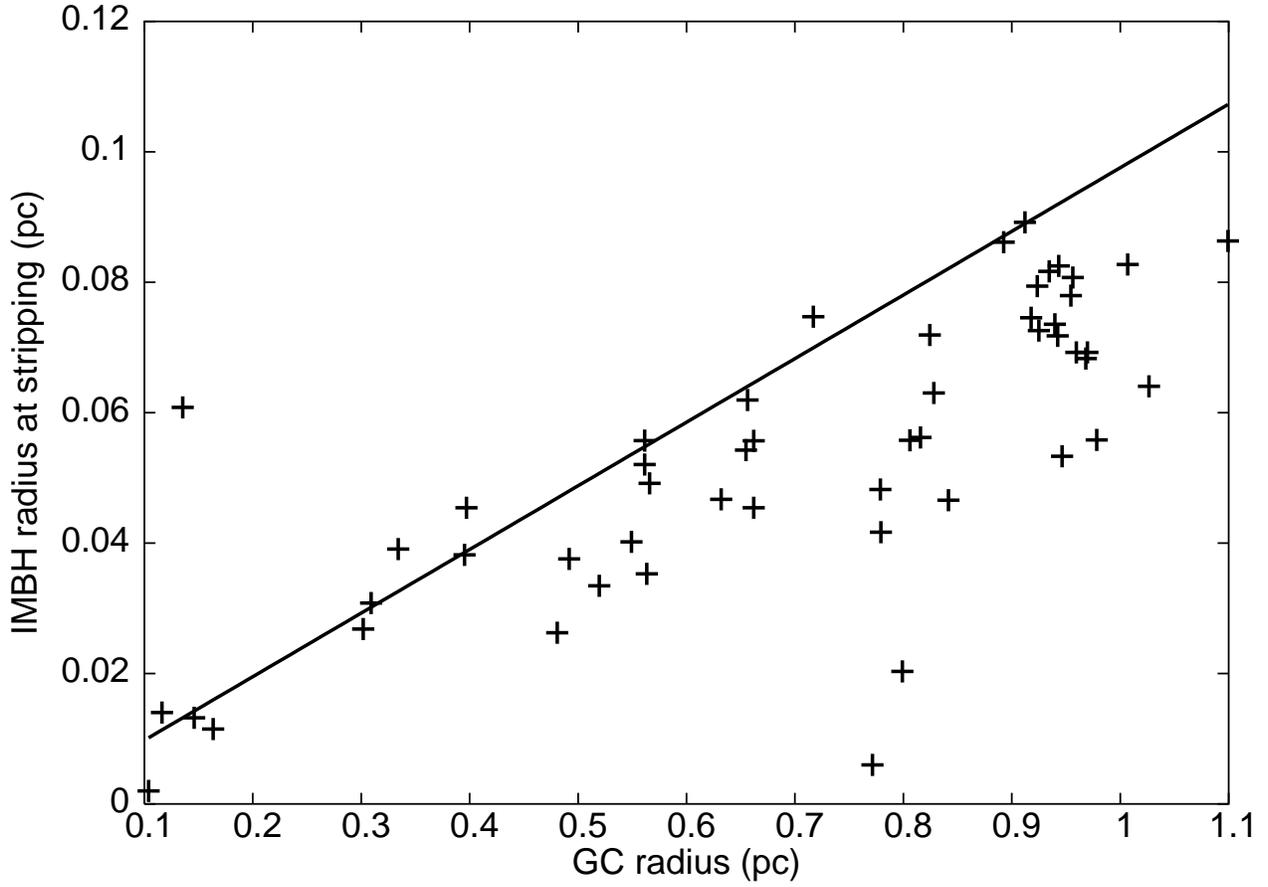}
\figcaption{
Comparison of Roche stripping and actual radii, for N-body case.  The abscissa is the star's radius from the SMBH, and the ordinate is the star's radius from the IMBH.  The standard Roche criterion provides only a rough upper limit on the actual stripping radii.
\label{fig:n_roche}}
\end{figure}

As in the three-body case, the standard escape energy criterion is better at determining actual stripping times, and the Roche criterion should be used solely as an order-of-magnitude estimate.  This point is emphasized again here, for the Roche criterion fails badly at determining stripping radii for many more stars in the N-body case.  Figure~\ref{fig:n_roche} shows that the standard Roche criterion $r\sim R(m/M)^{1/3}$ is an averaged upper limit in this environment.  Depending on the particular configuration, tidal forces may pull the star further or closer to the IMBH or SMBH, causing a scatter about the linear trend.  Indeed, two-body interactions are responsible for this deviation, since in the dense environment of a cluster core, star-star interactions force stellar orbits to slightly deviate from what would be expected in the three-body problem.    

\begin{figure}
\plotone{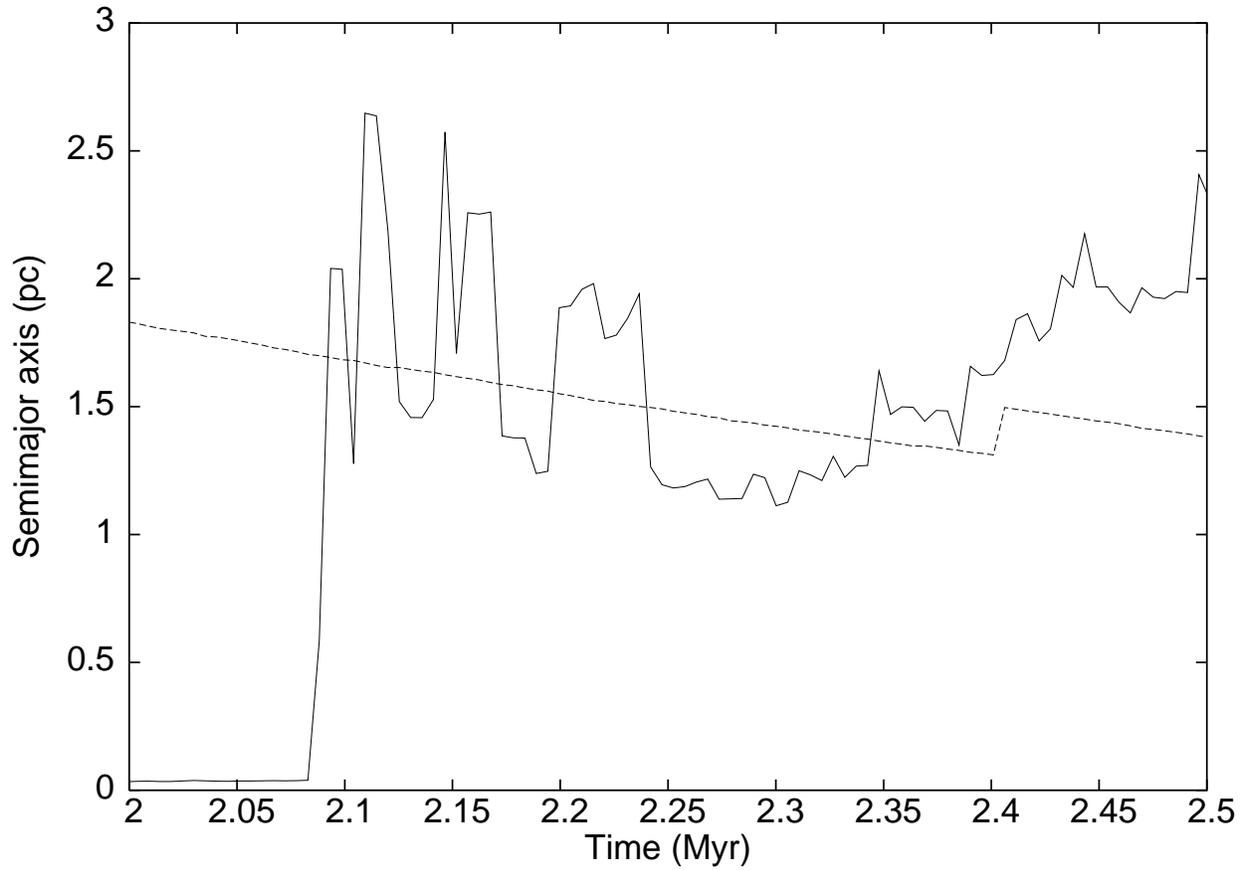}
\figcaption{The interaction of a star for the period following its loss from the cluster, in semimajor axis over time.  The dotted line is the IMBH and the jagged line is the star.  The star suffers several interactions with the IMBH while it is in a series of orbital resonances with it.  Eventually, the commensurability of the orbits is destroyed by subsequent interactions.  The jump discontinuity of the IMBH orbit is caused by a strong three-body interaction which resulted in a particle ejection.
\label{fig:strip}}
\end{figure}

Upon stripping, stars can be boosted into near resonant orbits with the IMBH.  It is instructive to follow the orbital evolution of such as star as it interacts with the IMBH potential well.  Figure~\ref{fig:strip} shows the interplay between a star and the IMBH for a few IMBH orbits.  Note that the star experiences wild oscillations in its semimajor axis due to the strong perturbing potential of the IMBH, but is not lost from the potential well for some time.  Other orbital elements will similarly be affected, and the result is that the memory of the stripping environment (i.e., the IMBH orbital elements) is made fuzzy by such encounters.  Note that this occurs in conjunction with the random walks due to small encounters discussed in \S~\ref{sec:scatter}.

\section{Discussion}
\label{sec:disc}

The goals of this study were twofold: to study the effectiveness with which an IMBH can transport a handful of stars deep into a SMBH potential well, and in the process, gain insight into the dynamics of young stars there.  The results of the three-body integrations suggest that the principal factors that influence the eventual deposition of stars are the IMBH orbital eccentricity and the inspiral speed.  Paumard et al. (2006) provide a review of particular features of the observations, and we compare our results to this.  
\begin{itemize}
\item The observations show two counter-rotating disks, oriented at large angles with respect to one another, with inner radii of about $0.05\pc$.  The cluster core simulations produce a single, pronounced disk of stars, but in no case is there any significant counter-rotating disk formed.  In order to explain the claimed multiple contemporaneous disks, one would need to invoke the almost simultaneous infall of two distinct clusters.
\item The disks are observed to have a well-defined inner radius, or edge.  The edge of the more populous disk (the clockwise disk) has an edge at 1$''$.  Such edges also emerge naturally from the cluster core simulations.  They result from the internal dynamical relaxation within the cluster, which sets up a
Bahcall-Wolf density cusp.  While individual stars may be sufficiently tightly bound to be transported to smaller radii, this edge represents the limit to which any significant cusp containing several stars may be transported. 
\item Paumard et al. (2006) find that both disks have surface density profiles that scale as $\Sigma \propto r^{-2}$.  Our simulations produce disks with surface density $\Sigma \propto r^{-0.75}$.  This result appears to be robust with respect to changes in the initial cluster density law.  We return to this below.
\item  There is an absence of stars on larger scales as might be expected from tidal stripping of less tightly bound material (e.g., Kim \& Morris 2003, Kim, Figer, Morris 2004).  Our simulations are devoted to the cluster core alone and do not address this directly.  Simulations of larger clusters suggest that this lack of observed massive stars could be a significant constraint on the cluster scenario (e.g., Portegies Zwart \& McMillan 2003). However, the observations are limited to massive stars and this constraint may be avoided if the massive stars in the cluster are initially centrally concentrated (G\"{u}rkan \& Rasio 2005), in which case they would be stripped only at small radii.
\item The disks are observed to have moderate thickness ($14\dgrs \pm 4\dgrs$ for the clockwise disk), which is well produced by our cluster core simulations.
\item Most of the stars in the clockwise disk are on low eccentricity orbits, while
 those in the counter-clockwise system are mostly on eccentric orbits.  Taken together, this is consistent with the two rings forming from entirely separate clusters, since the stripped disk stars tend to have eccentricities similar to those of the parent cluster orbit.  The fact that the more eccentric orbits lie further out is also consistent with the ability of circular IMBH orbits to transport stars deeper into the potential well.
\item Paumard et al. (2006) confirm earlier claims that the IRS13E 'clump' corresponds to a real overdensity.  Such core remnants emerge naturally from our simulations as well.  Further, our simulations mirror the observations in that such associations reside in the tidal tail of stripped stars.  

\end{itemize}

Thus, the cluster core simulations demonstrate an encouraging ability to reproduce several of the principal dynamical features of the observed disks.  In particular, the observed inner edge appears naturally and the presence of dynamically long-lived clumps appears generic, although the endurance of such clumps is constrained by high IMBH mass and low eccentricity.  Furthermore, we find that the cluster scenario naturally produces the observed thickness of the disks and results in similar eccentricities amongst stars in a given disk.  Nevertheless, there are several observed features which remain elusive. In no case do we produce a significant second disk -- if both disks are the result of cluster infall, then there must have been two separate clusters.  

Furthermore, the resulting surface density profile appears both robust and at odds with the observations.  It is a consequence of the internal dynamics of the cluster, and not a result of initial conditions specific to any particular formation scenario.  Internal dynamical relaxation leads to a cluster density profile given by the Bahcall \& Wolf (1976) law.  Tidal stripping of this cusp results in remnant disks with inevitable surface densities of $\Sigma \propto r^{-0.75}$.  The same internal density profile also explains the inner edge we observe. The number of surviving cluster stars, assuming a fixed Bahcall-Wolf density profile, scales as $N \propto r^{5/4}$. Thus, if the cluster core contained 100 stars at 1~pc, the radius at which a single star remains bound to the IMBH is $r \sim (1/100)^{4/5} \sim 0.025$~pc.  Thus, our simulations show that the gross features of the resulting population can be understood in large part by approximating the internal structure of the cluster as a relaxed Bahcall-Wolf cusp and treating the stripping with a simple Roche lobe criterion.

In addition to the disk stars, there is also the cluster of B-type stars close to the SMBH, the S-star cluster.  While our simulations do suggest that an IMBH can carry one or two stars very deep into the potential well, in no case do we find transport of a sufficient number of stars to explain the S-stars as the remnant of a single inspiral.  Further, the observed S-stars have varying semimajor axes and eccentricities.  These simulations show that at the end of its inspiral, the IMBH orbit has circularized, and stars acquire orbital parameters similar to that of the IMBH upon stripping.  This implies that the large variation in the orbital parameters of the S-stars cannot be accounted for simply by their deposition by a cluster infall.  Levin \& Beloborodov (2003) argue that Lens-Thirring precession may account for some randomization of the orbits, but this can only be effective for stars with initially small semimajor axes and large eccentricities, like SO-2; otherwise, the precession timescales are far too long, compared to the stellar lifetimes.  Infalling clusters likely do not survive to small radii, and are efficiently disrupted by strong peribothric tidal stresses when in highly eccentric initial orbits.  Thus, the origin and subsequent evolution of the S-stars remains unclear from these simulations, although a suite of scattering experiments is currently underway which may provide some insight.

Finally, the inclusion of a dark-mass cusp, perhaps representing a population of small black holes (Miralda-Escud\'e \& Gould 2000), significantly decreased the inspiral time of the IMBH.  This somewhat loosens the constraints on the infalling cluster scenario's ability to explain the peculiar dynamical environment at the Galactic Center, because ultimately, the scenario is constrained by the ages of the OB stars observed in the inner few arcseconds.  Consequently, given such a cusp, the delivery of very young OB stars to the Galactic Center can occur over a longer range of timescales than previously considered, since their $\sim 10\Myr$ measured ages allow them to have spent longer times outside the central parsec.  The outer portion of the inspiral thus may proceed more slowly, allowing the cluster time to further relax and mass-segregate.

\subsection{Comparison to other work}
A recent paper by Levin, Wu, \& Thommes (2005) reports on the simulations of a three-body problem similar to that treated in \S~\ref{sec:3bresults}.  They investigate the problem using a symplectic integrator in extended phase space, with an ad hoc treatment of close-encounters, which are the bane of many symplectic algorithms.  The IMBH orbit analytically decays, and all stars are set to be massless with the same initial Jacobi radius.  Further, their choice of inspiral timescale is in the range $1000-10000$ IMBH orbits.  They find that for circular IMBH inspirals, stars can achieve significant eccentricities but only low inclinations, even for slower inspiral.  These results are mirrored by those reported in \S~\ref{sec:3bresults}.  For eccentric inspirals, their results show two groups of stars, one in a thin disk of half-opening angle $10^{\circ}$, and the other with inclinations of $10^{\circ}<i<180^{\circ}$, but with randomized orbital parameters.  The former, but not the latter, are reproduced here as well.  Two points of departure may be their significantly longer inspiral times (roughly an order-of-magnitude) and the fact that their IMBH eccentricity is not allowed to evolve.  The first point allows more scattering events to occur, because the presence of the IMBH in adiabatically similar orbits will tend to produce more close encounters with stars recently stripped from the cluster into similar orbits.  This causes additional perturbations to orbital elements, lengthening their respective random walks and thus possibly increasing their net change.  

As for the second point, if the IMBH eccentricity is not allowed to vary, one is dismissing the true nature of strong interactions that may occur, and in particular, the often violent recoil of the IMBH due to ejections.  For an IMBH-star binary, an ejection may occur when an intruder makes a close passage to the binary, with kinetic energy exceeding the binary's binding energy.  Simple energy and momentum conservation arguments show that for typical systems simulated in this paper ($v_{BH}\sim 100 km/s$, $a_b\sim 10^{-4}\pc$), the perturbation $\delta{v}$ of the IMBH during such an event can be of order several percent of its orbital speed $v$.  The change in energy is linearly proportional to the change in semimajor axis, $v\delta{v} \sim a^{-2}\delta{a}$, and the eccentricity is affected similarly.  As a further consequence, the rest of the stars bound to the IMBH must respond to its perturbed motion by altering their orbits, and it is possible that any of these stars, perhaps marginally bound to the IMBH, will be lost ultimately due to the destruction of the original binary.  Thus, although the mass ratio in this problem is large, some important dynamics are missed in the test particle approximation. 

Given reasonable timescales, stars typically experience thousands of weak perturbations during a simulation, but only $\sim 20$ strong deflections.  Since these strong variations will dominate the net change, an analytic prescription for the IMBH orbit may yield incorrect results.  Comparison in further detail is probably superfluous, as a detailed discussion of the overall feasibility of the scenario requires consideration of the internal dynamics as well, neglected in any three-body treatment and the motivation for our N-body studies in \S~\ref{sec:nbody}.

A number of papers have supported the idea that infalling star clusters are responsible for some, if not all, of the kinematic structures at the Galactic center, but not without exception. In particular, Paumard et al. (2006) argue that IRS13E is an overdensity, potentially bound by the presence of an IMBH.  Sch\"odel et al. (2005), on the other hand, argue that the velocity dispersion of the IRS13E stars requires too large an IMBH mass and that there is no additional observational evidence (such as X-ray flares) to support the presence of an IMBH. Nevertheless, our calculations demonstrate that such configurations are dynamically possible, as tightly bound groups can be transported to where comoving groups like IRS13E are found.

A recent series of papers by Nayakshin and collaborators (Nayakshin 2005, Nayakshin et al. 2005, Nayakshin \& Sunyaev 2005, Nayakshin \& Cuadra 2005) also argues against the cluster model, in favor of an AGN-like accretion disk scenario.  Nayakshin \& Sunyaev (2005) note that the lack of hard X-ray emission from the GC region shows that there are few young, intermediate mass stars there, as might be expected from stars stripped from the putative inspiralling cluster. The degree to which this datum applies to the cluster scenario is unclear, as it depends on uncertain assumptions regarding the final mass function of cluster stars (which may very well be top-heavy at the start and further biased by the stellar merging that is an integral part of the IMBH formation scenario) as well as the final fraction of cluster mass that ends up in the IMBH.  Indeed, the lack of young stars is an important constraint on both cluster and disk scenarios, suggesting that the young star mass function is top-heavy, regardless of the kinematic origin of the population. The lack of any evidence for a stripped stellar population at large radii (Paumard et al 2006) is also an important constraint, but limited at present by the restricted area coverage of the observations.

Portegies Zwart et al. (2006) present results of large cluster inspiral simulations, from which they conclude that comoving groups can occur as IMBHs, formed through merger runaways, carry cluster remnants into the potential well.  The present simulations confirm that this occurs, although our requisite masses for transporting multiple stars deep into the well is somewhat above their canonical $1000 M_{\odot}$.  They also claim that several IMBH will be in the inner few milliparsecs at any given time.  This would provide a possibly efficient randomization mechanism for stellar orbits there, as well as a truncation mechanism for stellar disks.  However, at this time there is no clear observational evidence for a single IMBH, much less many; further, constraints on the Keplerian S-star orbits might rule out this possibility  (Ghez et al. 2005).

\subsection{Outlook}

In conclusion, we find that the simplest version of the IMBH+infalling cluster scenario can naturally produce several of the dynamical features of the population of young stars at the Galactic center, such as the disk thickness, the apparent inner edge, and the occurrence of dynamically long-lived clumps.  The  enthusiasm resulting from this agreement is, however, tempered somewhat by a few discrepancies that remain.  Most important of these are the differences between the observed and model surface density profiles and the apparent inability to transport a significant population of stars close enough to explain the S-star population.  Whether these represent the failure of the scenario as a whole, limitations of the numerical treatment, or simply an incompleteness of the model remains to be seen.

\acknowledgements
SB thanks the members of the UCLA Galactic Center group for useful comments on earlier drafts of this paper, and Sverre Aarseth for advice on the intricacies of NBODY6.

\clearpage

\clearpage
\clearpage

\clearpage
\clearpage

\clearpage
\clearpage

\clearpage
\begin{figure}
\plotone{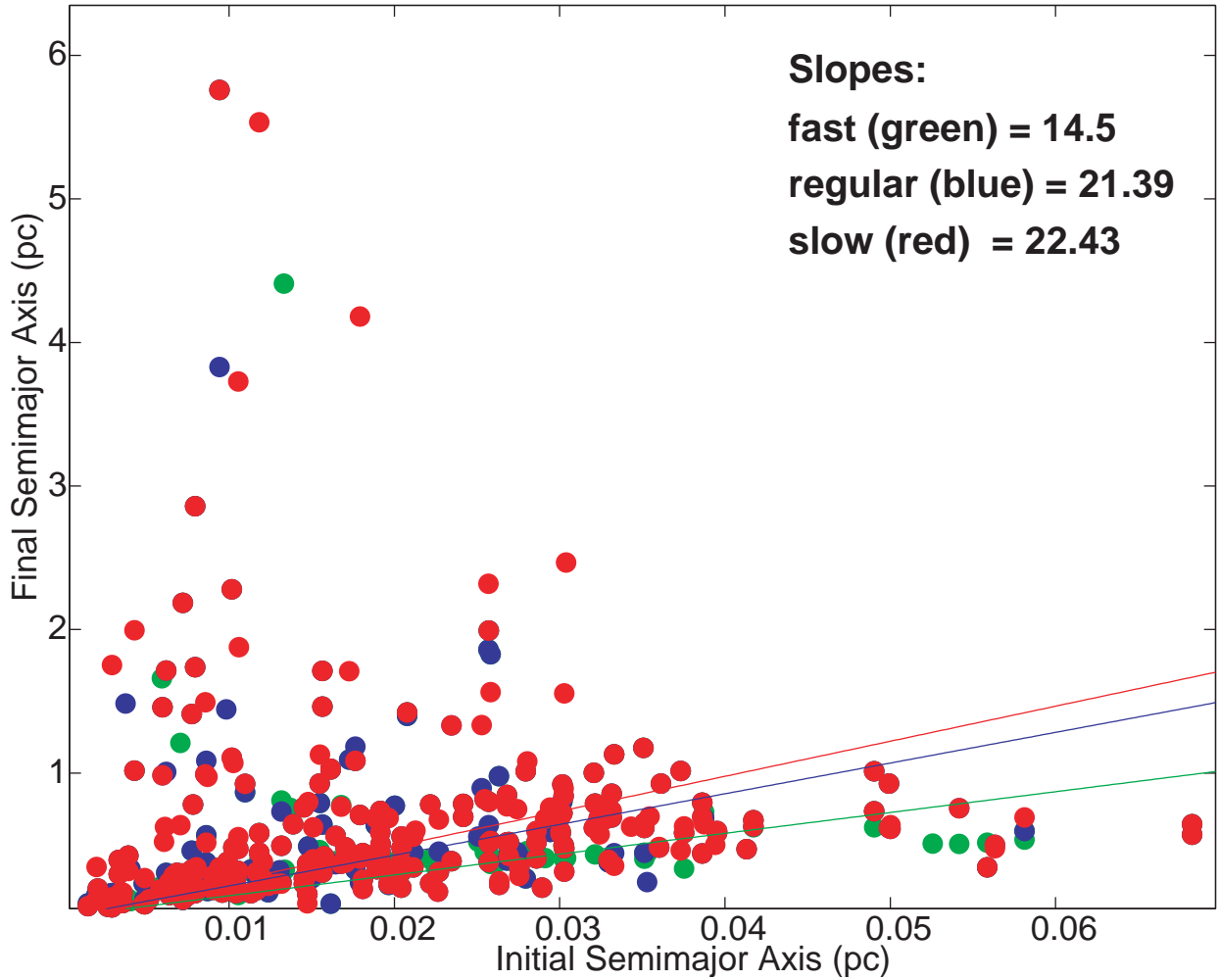}
\figcaption{
Comparison of final and initial semimajor axes, for initial IMBH eccentricity $e=0.8$.  Note that there are more stars in final orbits of semimajor axes for slow inspirals.  This trend is sustained also for $e=0.0$ inspirals (not shown).  Initial values are for stars in IMBH-centric orbits, while final values specify SMBH-centric values. 
\label{fig:blah}
}
\end{figure}
\clearpage

\clearpage
\clearpage

\clearpage
\clearpage

\clearpage
\clearpage

\clearpage
\input{tab1.tex}
\clearpage
\clearpage

\clearpage
\clearpage

\clearpage
\clearpage

\clearpage
\clearpage

\clearpage

\clearpage

\clearpage
\clearpage

\clearpage
\clearpage

\clearpage
\clearpage

\clearpage
\clearpage

\clearpage
\clearpage

\clearpage
\clearpage

\clearpage
\clearpage

\end{document}

%% file: tab1.tex
\begin{deluxetable}{cccccc}
\tablewidth{0pt}
\tablecolumns{6}
\tablecaption{Smallest IMBH semimajor axis with retention at least 6 stars forming comoving groups.\label{tbl:com}}
\tablehead{
  \multicolumn{3}{c}{Without Cusp} & \multicolumn{3}{c}{With Cusp} \\
  \cline{1-3} \cline{4-6} \\
  \colhead{$M_{IMBH}$} & \colhead{$e$} & \colhead{$a_6 (pc)$} &
  \colhead{$M_{IMBH}$} & \colhead{$e$} & \colhead{$a_6 (pc)$}}
\startdata
5540 & 0    & 0.131 & 5426 & 0    & 0.129 \\
5426 & 0.25 & 0.134 & 4062 & 0    & 0.131 \\
5540 & 0.25 & 0.147 & 5540 & 0.25 & 0.146 \\
5512 & 0.25 & 0.152 & 5540 & 0    & 0.152 \\
4134 & 0.25 & 0.158 & 4155 & 0.25 & 0.159 \\
5512 & 0    & 0.170 & 5512 & 0.25 & 0.166 \\
4155 & 0    & 0.170 & 4155 & 0    & 0.169 \\
5426 & 0.5  & 0.185 & 5426 & 0.5  & 0.200 \\
4062 & 0    & 0.200 & 4134 & 0    & 0.213 \\
5426 & 0    & 0.204 & 5426 & 0.25 & 0.214 \\
5540 & 0.5  & 0.228 & 4062 & 0.5  & 0.217 \\
4134 & 0    & 0.249 & 4062 & 0.25 & 0.234 \\
4062 & 0.25 & 0.266 & 4155 & 0.5  & 0.239 \\
4155 & 0.25 & 0.269 & 5512 & 0.5  & 0.259 \\
4062 & 0.5  & 0.275 & 5540 & 0.5  & 0.269 \\
4134 & 0.5  & 0.294 & 2031 & 0.5  & 0.278 \\
5512 & 0.5  & 0.294 & 2078 & 0    & 0.281 \\
2067 & 0.5  & 0.303 & 5512 & 0    & 0.287 \\
4155 & 0.5  & 0.313 & 2078 & 0.5  & 0.302 \\
2078 & 0.5  & 0.322 & 4134 & 0.5  & 0.310 \\
1357 & 0.25 & 0.335 & 2067 & 0    & 0.320 \\
2031 & 0.25 & 0.335 & 2031 & 0.25 & 0.326 \\
1385 & 0.5  & 0.349 & 1378 & 0.5  & 0.342 \\
2031 & 0.5  & 0.350 & 2031 & 0.5  & 0.345 \\
2067 & 0.25 & 0.378 & 2067 & 0.25 & 0.348 \\
2078 & 0    & 0.399 & 4134 & 0.25 & 0.363 \\
2031 & 0    & 0.410 & 1378 & 0    & 0.390 \\
1357 & 0.5  & 0.413 & 1039 & 0.5  & 0.419 \\
1039 & 0.5  & 0.443 & 1015 & 0.25 & 0.427 \\
2067 & 0    & 0.468 & 1357 & 0.25 & 0.443 \\
1378 & 0    & 0.487 & 2078 & 0.25 & 0.447 \\
2078 & 0.25 & 0.510 & 1367 & 0.25 & 0.469 \\
1015 & 0.5  & 0.518 & 1015 & 0.25 & 0.479 \\
1357 & 0    & 0.525 & 1378 & 0.25 & 0.481 \\
1378 & 0.25 & 0.528 & 2031 & 0    & 0.518 \\
1385 & 0.25 & 0.530 & 1357 & 0    & 0.550 \\
1034 & 0.25 & 0.559 & 1034 & 0    & 0.623 \\
\enddata
\end{deluxetable}

%% file: ms.bbl
\begin{thebibliography}{}
\bibitem[Aarseth(2003)]{Aarseth03} Aarseth, S.J. 2003, Gravitational N-body simulations (Cambridge: Cambridge University Press)
\bibitem[Bahcall \& Wolf(1976)]{BW76} Bahcall, J. N. \& Wolf, R. A.  1976, \apj, 209, 214
\bibitem[Baumgardt et al.(2004)]{Baumgardt04} Baumgardt, H., Makino, J., Ebisuzaki, T.  2004, \apj, 613, 1133
\bibitem[Binney \& Tremaine(1987)]{Binney87} Binney, J. and Tremaine, S. 1987, Galactic Dynamics (Princeton: Princeton University Press)
\bibitem[Chandrasekhar \& von Neumann(1943)]{Chandra43} Chandrasekhar, S., \& Von Neumann, J.  1943, \apj, 97, 1
\bibitem[Del Popolo, A. \& Gambera, M.(1999)]{Delpopolo99} Del Popolo, A. \& Gambera, A.  1999,  A\&A, 342, 34
\bibitem[Eisenhauer et al.(2005)]{Eisen05} Eisenhauer, F., et al.  2005, \apj, 628, 246
\bibitem[Gerhard(2001)]{Gerhard01} Gerhard, O.  2001, \apj, 546, L39
\bibitem[Genzel et al.(2000)]{Genzel00} Genzel, R., Pichon, C., Eckart, A., Gerhard, O. E. \& Ott, T.  2000, \mnras, 317, 348
\bibitem[Genzel et al.(2003)]{Genzel03} Genzel, R., et al.  2003, \apj, 594, 812
\bibitem[Ghez et al.(2003)]{Ghez03} Ghez, A. M. et al.  2003, \apj, 586, L127
\bibitem[Ghez et al.(2005)]{Ghez05} Ghez, A.M. et al.  2005, \apj, 620, 744
\bibitem[Goodman(2003)]{Goodman} Goodman, J.  2003, \mnras, 339, 937
\bibitem[G\"{u}rkan \& Rasio(2005)]{Ato05} G\"urkan, A. \& Rasio, F.  2005, \apj, 628, 236
\bibitem[Hansen \& Milosavljevi\a'c(2003)]{Hansen03} Hansen, B. and Milosavljevi\a'c, M.  2003, \apj, 593, L77
\bibitem[Kim \& Morris(2003)]{Kim03} Kim, S., Morris, M.  2003, \apj, 597, 312
\bibitem[Kim et al.(2004)]{Kim04} Kim, S., Figer, D., and Morris, M.  2004, \apj, 607, L123
\bibitem[Kolykhalov \& Sunyaev(1980)]{Koly80} Kolykhalov, P. \& Sunyaev, R. 1980, SvAL, 6, 357
\bibitem[Kroupa et al.(1993)]{Kroupa93} Kroupa, P., Tout, C.A., Gilmore, G.  1993, \mnras, 262, 545.
\bibitem[Levin \& Beloborodov(2003)]{Beloborodov03} Levin, Y., and Beloborodov, S.  2003, \apj, 590, 33L
\bibitem[Levin, Wu, Thommes(2005)]{Levin05} Levin, Y., Wu, A., Thommes, E.  2005, \apj, 635, 341
\bibitem[Lu et al.(2005)]{Lu05} Lu, J.R., Ghez, A.M.; Hornstein, S.D.; Morris, M.; Becklin, E.E.  2005, \apjl, 625, 51L
\bibitem[Maillard et al.(2004)]{Maillard04} Maillard, J.P., Paumard, T., Stolovy, S.R., and Rigaut, F.  2004, A\&A, 423, 155
\bibitem[McMillan \& Portegies Zwart(2003)]{McMillan03} McMillan, S.L. and Portegies Zwart, S.F.  2003, \apj, 596, 314
\bibitem[Miralda-Escud\a'e \& Gould(2000)]{Miralda00} Miralda-Ecud\a'e, J. \&  Gould, M.  2000, \apj, 545, 847
\bibitem[Morris \& Serabyn(1996)]{Morris96} Morris, M. \& Serabyn, E. 1996, ARA\&A, 24, 345 
\bibitem[Murray \& Dermott(1999)]{Murray99} Murray, C.D. and Dermott, S.F.  1999, Solar System Dynamics (Cambridge: Cambridge University Press)
\bibitem[Nakano \& Makino(1999)]{Nakano99} Nakano, T., and Makino, J.  1999, \apj, 525, L77
\bibitem[Nayakshin \& Cuadra(2004)]{Nay04} Nayakshin, S. \& Cuadra, J.  2004, A\&A, 437, 437
\bibitem[Nayakshin et al.(2005)]{Nayetal05} Nayakshin, S., Dehnen, W., Cuadra, J., Genzel, R.  2005, preprint (astro-ph/0511830)
\bibitem[Nayakshin \& Sunyaev(2005]{NayS05} Nayakshin, S. \& Sunyaev, R.  2005, \mnras, 364L, 23
\bibitem[Nayakshin(2005)]{Nay05} Nayakshin, S. \& Sunyaev, R.  2005, preprint (astro-ph/0507687)
\bibitem[Portegies Zwart et al.(2005)]{PZ05} Portegies Zwart, S., Baumgardt, H., McMillan, S., Makino, J., Hut, P.  2005 preprint (astro-ph/0511397)
\bibitem[Reid \& Brunthaler2004]{RB04} Reid, M.J. \& Brunthaler, A.  2004, \apj, 616, 872
\bibitem[Sanders(1992)]{Sanders92} Sanders, R.H.  1992, Nature, 359, 131
\bibitem[Sanders(1998)]{Sanders98} Sanders, R.H.  1998, MNRAS, 294, 35
\bibitem[Shlosman \& Begelman(1989)]{Sh89} Shlosman, I. \& Begelman, M. 1989, \apj, 341, 685
\bibitem[Sch\"odel(2005)]{Schodel05} Sch\"odel, R., Eckart, A., Iserlohe, C., Genzel, R., Ott, T.  2005, \apj, 625, 111L
\bibitem[Spinnato(2003)]{Spinnato03} Spinnato, P.F., Fellhauer, M. and Portegies Zwart, S.F.  2003, \mnras, 22, 32
\bibitem[Spitzer(1987)]{Spitzer87} Spitzer, L.  1987, Dynamical evolution of globular clusters (Princeton:Princeton University Press)
\bibitem[Yu \& Tremaine(2001)]{YuT01} Yu, Q. \& Tremaine, S.  2001, A\&A, 121, 1736
\end{thebibliography}
